\documentclass[a4paper,11pt]{article}
\pdfoutput=1 

\usepackage{jheppub} 

\usepackage[T1]{fontenc} 
\usepackage{dcolumn}
\newcolumntype{d}{D{.}{.}{2.3}}
\usepackage{xspace}
\usepackage[export]{adjustbox}
\usepackage{subcaption}
\newcommand{\as}{\ensuremath{\alpha_s}\xspace}
\newcommand{\di}{\ensuremath{\mathrm{d}}\xspace}
\newcommand{\mjj}{\ensuremath{m_{j_1 j_2}}\xspace}
\newcommand{\Ht}{\ensuremath{H_T}\xspace}
\newcommand{\Htpartonic}{\ensuremath{\hat{H}_T}\xspace}
\newcommand{\htt}{\ensuremath{\hat H_T/2}\xspace}
\newcommand{\LL}{\ensuremath{LL}\xspace}
\newcommand{\LLp}{\ensuremath{LL^+}\xspace}

\newcommand{\Egt}{\ensuremath{E_\gamma^T}\xspace}
\newcommand{\HEJ}{HEJ\xspace}
\newcommand{\HighEJ}{High Energy Jets\xspace}
\newcommand{\murf}{\ensuremath{\mu_{\scriptscriptstyle \mathrm{R},\scriptscriptstyle \mathrm{F}}}}
\newcommand{\fbi}{\ensuremath{\mathrm{fb}^{-1}}\xspace}

\hypersetup{pdftitle=Stable perturbative predictions for isolated photon production with a jet pair at large mj1j2,pdfdisplaydoctitle}
\title{\boldmath Stable perturbative predictions for isolated photon production with a jet pair at large $m_{j_1 j_2}$\unboldmath}
\author[a,b]{Jeppe R.~Andersen,}
\author[c,d]{Andreas Maier,}
\author[a]{Malina Rosca,}
\author[e]{Giacomo Ruisi}

\affiliation[a]{Institute for Particle Physics Phenomenology, University
  of Durham, South Road, Durham DH1 3LE, UK}
\affiliation[b]{Theoretical Physics Department, CERN, 1211 Geneva 23, Switzerland}
\affiliation[c]{Institut de F{\'i}sica d'Altes Energies (IFAE), The Barcelona Institute of Science and Technology, Campus UAB, 08193 Bellaterra (Barcelona), Spain}
\affiliation[d]{Grup de F{\'i}sica Te\`orica, Dept. F{\'i}sica,
Universitat Aut\`onoma de Barcelona, E-08193 Bellaterra, Barcelona, Spain}
\affiliation[e]{Deutsches Elektronen-Synchrotron DESY, Platanenallee 6, 15738 Zeuthen, Germany}

\emailAdd{jeppe.andersen@durham.ac.uk}
\emailAdd{amaier@ifae.es}
\emailAdd{malina.rosca@durham.ac.uk}
\emailAdd{giacomo.ruisi@desy.de}

\preprint{CERN-TH-2025-120, DESY-25-109, IPPP/25/35}

\abstract{We present the first calculation of the high-energy corrections to the
  process of isolated photon production in association with two jets. These corrections both stabilise the perturbative behaviour in $\mjj$ and lead to a
  significantly improved description of data from a recent ATLAS
  measurement.}

\begin{document}
\maketitle
\flushbottom

\section{Introduction}
\label{sec:intro}
The impressive efforts of the LHC experiments in achieving increasingly precise measurements of the cross sections of ever rarer Standard Model processes~\cite{ATLAS:2024cgh} is met by a Herculean effort of achieving similarly precise fixed order predictions from first principles, such that the measurement can be interpreted in terms of the underlying theory.
This pursuit has resulted in agreement between prediction and measurement for inclusive processes measured at the LHC which have cross sections spanning more than 9 orders of magnitude.
This is true in the region where the notion of a single hard perturbative scale should apply.

Kinematic distributions rather than just the cross sections are often needed when searching for possible signals of new interactions and physics.
An example of such a study is a recent measurement from ATLAS of the production of missing transverse momentum and that of a photon, both in association with either at least a single jet or at least two jets~\cite{ATLAS:2024vqf}.
This followed an earlier study~\cite{ATLAS:2019iaa} of the same process.

The theory predictions confronted with these measurement were met with several problems in describing specifically the normalisation and the kinematic distribution in \mjj, the invariant mass between the two hardest jets.
In fact, the description of the recent ATLAS measurement~\cite{ATLAS:2024vqf} was so poor that the whole dataset on $\gamma$+dijets was excluded from the analysis on BSM exclusion.\footnote{The description of the \mjj spectrum for $Z$+dijets is also poor.}
The fixed order predictions have large scale dependence and large corrections from NLO to NNLO~\cite{Badger:2023mgf} and slow numerical convergence for large \mjj. In fact, the NNLO results converge only up to \mjj around 1TeV, whereas the recent measurement by ATLAS~\cite{ATLAS:2024vqf} extends to 8.2TeV.
The shower-based merging of NLO-matched samples~\cite{Hoeche:2012yf}, while numerically stable, fails to describe the distribution at large \mjj.

In the current study we will calculate the predictions for the photon+dijets measurements in reference~\cite{ATLAS:2024vqf} and demonstrate that the all-order resummation of logarithms relevant at large \mjj obtains a stable, precise and accurate prediction of data, where fixed next-to-leading order and other predictions struggle to describe the distribution.

Section~\ref{sec:measurement} will detail the relevant components of the measurement for this study.
Section~\ref{sec:fixedorder} explains the challenges posed to fixed order for the current measurement.
The resummation leading to perturbative stability is introduced in section~\ref{sec:corlarges}, and the comparison to data is presented in section~\ref{sec:CompData}.
Finally, the conclusions are presented in section~\ref{sec:Conclusions}.
The appendices~\ref{sec:comp-high-energy}--\ref{sec:mjj} contain further discussions of the resummation and of fixed order results using an alternative choice of factorisation and renormalisation scale.

\section{ATLAS 13 TeV Measurements of Photon+Dijets  and the Status of Predictions}
\label{sec:measurement}
The production of prompt photons in association with jets at the LHC energy of 13TeV has been studied in two ATLAS papers with an integrated luminosity of 36.1~\fbi~\cite{ATLAS:2019iaa} and 140~\fbi~\cite{ATLAS:2024vqf} focusing on a very inclusive and a very exclusive measurement respectively.
The comparison with NNLO\footnote{NNLO in the QCD contribution, see later for a discussion of the electro-weak component.}~\cite{Badger:2023mgf} focuses on the ``direct enriched'' sample of the inclusive measurement with the cuts listed in table~\ref{tab:inclusive}.
\begin{table}
  \centering
  \begin{tabular}{ll}\hline
    Photon rapidity & $|y|\le 1.37$ or $1.56\le|y|\le 2.37$\\
    Photon energy & $E_\gamma^T>150$~GeV\\
    Jet transverse momentum & $p_T^{\mathrm{jet}}$>100~GeV\\
    Jet rapidity & $|y^\mathrm{jet}|<2.5$\\
    Minimum number of jets & 2\\
    Jet-photon separation & $\Delta R^{\gamma\mathrm{-jet}}>0.8$\\
    Phase space cut & $E_T^\gamma>  p_T^{j_1}$\\
    Jet definition & anti-$k_t$, $R=0.4$\\\hline
  \end{tabular}
  \caption{Cuts for the 36.1~\fbi sample analysed by ATLAS~\cite{ATLAS:2019iaa}. Additional cuts were imposed vetoing hadronic activity in an isolation cone around the photon.}
  \label{tab:inclusive}
\end{table}
The hard cut on the photon and the requirement on photon isolation ensure this sample is dominated by contributions where the photon is produced directly from the hard process, rather than through the fragmentation of the produced hard partons.
This study includes only central jets ($|y^\mathrm{jet}|<2.5$), and the cuts on the transverse energies and momenta of jets and photons are slightly different than for the measurement to be studied later.
Most importantly, this study is \emph{inclusive} in the number of jets (no central jet veto as in~\cite{ATLAS:2024vqf}).
However, we can draw some conclusions which will be of relevance for the detailed study of the more recent measurement~\cite{ATLAS:2024vqf} to follow.
The first conclusion already mentioned in~\cite{ATLAS:2019iaa} is that for this process even the NLO-matched and multi-jet merged approach detailed in~\cite{Hoeche:2012yf} describes data only for $\mjj<1$TeV (see figure~9 of~\cite{ATLAS:2019iaa}).

The second set of observations of relevance for the later studies are from the presentation of the first full NNLO calculation~\cite{Badger:2023mgf} for this process.
First, we note (figure~6 of reference~\cite{Badger:2023mgf}) that pure NLO and MEPS@NLO have very different shapes in \mjj even though this is an inclusive two-jet observable.
This is of course permitted, even if the calculation share the same $\alpha_s^2\alpha$ and $\alpha_s^3\alpha$ terms.
Secondly, the NNLO corrections can lower the normalisation of the cross section obtained from NLO by roughly 15\%, depending on the scale choice.
The NNLO calculation~\cite{Badger:2023mgf} studied the similar scale choices of $E_\perp^\gamma$ and $H_T$ (the scalar sum of the transverse momenta of the photon and of the two hardest jets).
It would be interesting to understand further whether this change in normalisation from NLO to NNLO is related to the logarithmic corrections from small-cone isolation~\cite{Catani:2013oma,Balsiger:2018ezi,Becher:2022rhu} or to super-leading logarithms and other higher-order effects~\cite{Becher:2023mtx,Boer:2024xzy,Becher:2024nqc,Boer:2024hzh,Forshaw:2006fk,Forshaw:2008cq,AngelesMartinez:2018cfz,Banfi:2021xzn,Becher:2024kmk}.
While the NLO calculation is numerically stable for this inclusive measurement across the range of \mjj studied up to 4TeV, the NNLO calculation displays large numerical variations in the bins beyond 1TeV.
The authors identify the cause of the lack of numerical convergence as being due to the large cancellations between the various components of the NNLO calculation~\cite{Badger:2023mgf}.

The recent ATLAS study~\cite{ATLAS:2024vqf} used 140\fbi of data to analyse photon production in association with dijets within the cuts listed in table~\ref{tab:cuts}.
\begin{table}
  \centering
  \begin{tabular}{ll}\hline
    Photon rapidity & $|y|\le 1.37$ or $1.52\le|y|\le 2.47$\\
    Leading photon $p_T$ & $>160$ GeV\\
    $p_T^\mathrm{recoil}$ & $>200$ GeV\\
    Leading jet $p_T$ & $>80$ GeV\\
    Sub-leading jet $p_T$ & $>50$ GeV\\
    Leading jet $|y|$ & $<4.4$\\
    Sub-leading jet $|y|$ & $<4.4$\\
    Dijet invariant mass \mjj & $>200$ GeV\\
    $|\Delta y_{j_1 j_2}|$ & $>1$\\
    Jets with rapidity in-between hardest two& None with $p_T>30$ GeV\\
    Jet definition & anti-$k_t$, $R=0.4$\\\hline
  \end{tabular}
  \caption{Cuts for the analysis reported in reference~\cite{ATLAS:2024vqf}. The analysis is implemented in Rivet~\cite{Bierlich:2024vqo}.}
  \label{tab:cuts}
\end{table}
A few differences in the cuts compared to the earlier ATLAS analysis are worth pointing out, since they will impact the perturbative predictions significantly.
While the cuts on the transverse momentum of the photon and of the jets are changed only slightly, the rapidities of the jets in reference~\cite{ATLAS:2024vqf} can extend to 4.4, allowing for much larger inter-jet rapidities and larger \mjj.
The most important difference though is the requirement of the absence of a third jet above 30~GeV with a rapidity in-between the two hardest jets, which are also required to be at least one unit of rapidity apart.
While the minimum rapidity separation requirement is modest, the rapidity veto removes the contribution from a third jet in-between the two hardest jets from the corresponding inclusive prediction.
It therefore introduces a perturbative effect proportional to $\alpha_s |\Delta y_{j_1j_2}|$ relative to a setup similar to the perturbative calculation~\cite{Badger:2023mgf}, which is relevant for the inclusive measurement~\cite{ATLAS:2019iaa}.

The following three kinematic distributions  were reported by ATLAS (see figure 12 of~\cite{ATLAS:2024vqf}):
\begin{enumerate}
\item $\di\sigma/\di p^\gamma_{T}$ in 11 bins $200\ \mathrm{GeV}<p^\gamma_{T}<2600\ \mathrm{GeV}$
\item $\di\sigma/\di \mjj$ in 14 bins for $200\ \mathrm{GeV}<\mjj<8200\ \mathrm{GeV}$
\item $\di\sigma/\di \Delta \phi_{j_1 j_2}$ in 20 bins for $-1< \Delta \phi_{j_1 j_2}/\pi<1$
\end{enumerate}
In line with the conclusions from the inclusive measurement in reference~\cite{ATLAS:2019iaa}, the prediction from MEPS@NLO~\cite{Hoeche:2012yf} is uniformly 15\%-20\% high for the predictions for $\di\sigma/\di p^\gamma_{T}$ and $\di\sigma/\di \Delta \phi_{j_1 j_2}$ of reference~\cite{ATLAS:2024vqf}, whereas even the shape of $\di\sigma/\di \mjj$ is very poorly described (see figure 12 of~\cite{ATLAS:2024vqf}) with a systematically increasing deviation from data for increasing \mjj.
This very poor Standard Model description of the spectrum meant that the data on photon+dijets had to be disregarded from the BSM search and exclusion study in reference~\cite{ATLAS:2024vqf}.

We will in the next section investigate the fixed order predictions, and in section \ref{sec:corlarges} demonstrate how stable predictions can be obtained using a purely perturbative approach.

\section{Investigating Fixed Order Predictions}
\label{sec:fixedorder}
The process $pp\to\gamma jj$ has at Born level both a QCD $\alpha_s^2\alpha$ and an electro-weak $\alpha^3$ component, and both are equally relevant at large \mjj.
Our primary focus will be on the QCD component, since the perturbative corrections to the electro-weak component are better behaved and controlled.
The interference between the QCD and electro-weak (EW) process will be suppressed by colour effects just as for the QCDxEW effects in $pp\to hjj$~\cite{Andersen:2007mp,Bredenstein:2008tm,Dixon:2009uk}.

We will start the perturbative investigation by calculating the QCD dominated contribution to $pp\to\gamma jj$ at $\alpha_s^2\alpha$ (Born level) and up to $\alpha_s^3\alpha$ (next-to-leading order), both using the conventional choice for renormalisation and factorisation scale of \htt (half the scalar sum of the transverse momenta of the final state particles).
A conventional 7-point scale variation with a factor of 2 is employed.
We will focus our attention solely on the distribution $\di\sigma/\di\mjj$, since it will cleanly illustrate the issues which arise.
The fixed order calculations of the QCD process are performed using Sherpa 2.2.16~\cite{Sherpa:2019gpd} and OpenLoops~\cite{Buccioni:2019sur}, with the analysis implemented in Rivet~4~\cite{Bierlich:2024vqo}, employing the anti-$k_t$~\cite{Cacciari:2008gp} jet-algorithm implemented in FastJet~\cite{Cacciari:2005hq}.
For all the predictions presented in this study we use the NNPDF3.1~\cite{NNPDF:2017mvq} set provided by LHAPDF~\cite{Buckley:2014ana}. We have checked that the difference to PDF4LHC21~\cite{PDF4LHCWorkingGroup:2022cjn} is insignificant at small \mjj and grows to 10\% in the last bin in the \mjj-distribution.

\begin{figure}[tb]
  \centering
  \includegraphics[width=0.8\textwidth]{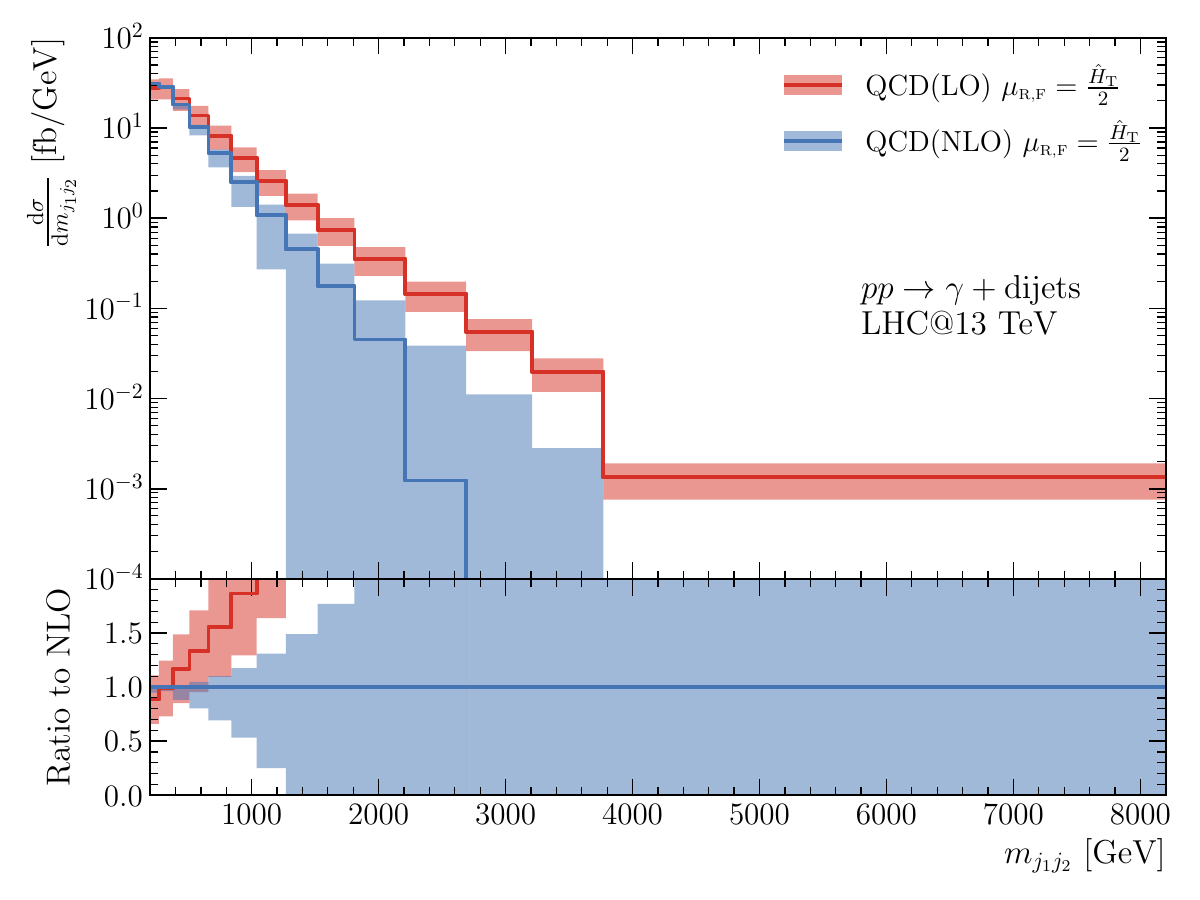}
  \caption{$\di\sigma/\di \mjj$ for LO and NLO for $pp\to \gamma jj$($\as^2 \alpha$) with central scale choice $\mu_R=\mu_F=\htt$ and a standard 7-point scale variation.}
  \label{fig:fo}
\end{figure}
Figure~\ref{fig:fo} shows the Born level prediction of $\di\sigma/\di\mjj$ including scale variation in red and the NLO prediction in blue.
While the Born level prediction remains physical for all \mjj, the central prediction at NLO turns small or even negative around 3TeV, and the width of the scale variation band increases uniformly with \mjj.
There is overlap between the scale variation bands at LO and NLO only for small $\mjj<800$GeV.

It may at first be hard to reconcile the stability found at NLO in reference~\cite{Badger:2023mgf} with the result in figure~\ref{fig:fo}.
However, several differences are worth highlighting. Firstly, the central jet veto in the current analysis has removed a large positive contribution of the form $\as\Delta y\sim\as\log(\mjj/p_t)$ from the inclusive setup in~\cite{ATLAS:2019iaa,Badger:2023mgf}.
Secondly, using the scale \Ht results in the upper scale variation limit, which stays positive even for the jet vetoed cross section up to the 4TeV limit relevant for the inclusive studies of~\cite{ATLAS:2019iaa,Badger:2023mgf}.
The cross section is of course completely determined by the behaviour in the first few bins, and the fact that its value turns negative or close to zero for $\mjj>4$TeV is a problem only for that particular distribution, not even for the other two distributions studied.

Why would we then choose \htt rather than \Htpartonic as the central scale for the calculations reported here and in later sections?
\htt is a standard scale choice used in the predictions for vector boson production in association with jets~\cite{Berger:2009ep,Berger:2010zx,Ita:2011wn,Bern:2013gka}.
Choosing \Htpartonic does not solve the problem, it just pushes both the explosion of the size of the scale variation and the negativity to larger \mjj.
As such, choosing any specific multiple of \Htpartonic to make LO and NLO similar would seem like a perturbative fine tuning rather than a useful choice for a true prediction.
Even the calculation at NNLO~\cite{Badger:2023mgf} presents two similar scale choices (\Ht and \Egt) which lead to different conclusions about the size of the NNLO corrections.\footnote{It would be interesting if an analysis as the one done in reference~\cite{Caola:2021mhb} could identify a reasonable dynamic scale across phase space.}
As mentioned in section~\ref{sec:measurement}, even the NNLO corrections~\cite{Badger:2023mgf} on the total cross section were reported to be up to 15\% compared to NLO.

In conclusion, we have identified issues of perturbative instability  for large \mjj for the next-to-leading order predictions relevant for the measurement in reference~\cite{ATLAS:2024vqf}.

\section{Establishing Perturbative Stability at Large Partonic Invariant Mass}
\label{sec:corlarges}
Having identified issues in the perturbative series at large \mjj, and noting that $\mjj<\sqrt s$ (such that large \mjj means large $\sqrt s$), it becomes interesting to investigate the predictions obtained within a perturbative approach which systematically addresses the perturbative logarithms arising in $s$.
These so-called high energy logarithms can be resummed using the BFKL~\cite{Fadin:1975cb,Kuraev:1976ge,Kuraev:1977fs,Balitsky:1978ic} equation.
However, this approach is \emph{inclusive} in any emission, and therefore not directly suitable for a measurement employing a central jet veto.

We will use here the framework of \HighEJ (\HEJ).
This framework captures the same logarithms from full QCD as BFKL, but the all-order resummation is constructed by an explicit sum over the (numerical) phase space integral over each multiplicity, and it is therefore possible to use standard jet-algorithms and analyses.
Indeed, the same packages mentioned in the discussion for the fixed-order calculations in section~\ref{sec:fixedorder} are used for the analyses of the results obtained with \HEJ.
Furthermore, the amplitudes in \HEJ differ in several ways from those of BFKL in retaining the gauge-invariance, crossing symmetry and Lorentz invariance of full QCD, see reference~\cite{Andersen:2020yax} for details.

This is the first study of photon production in association with dijets with \HEJ, and several new components had to be calculated.
We include selected sub-leading logarithmic corrections, denoting the accuracy as \LLp.
The difference between \LL and \LLp amounts to roughly 20\% almost uniformly across \mjj.
We match the cross section multiplicatively to NLO, observing a small matching correction of 0.5\% for the central scale choice and a surprisingly good agreement for $\di\sigma/\di\mjj$ between NLO and the NLO-truncated resummation.
Further details can be found in appendices~\ref{sec:comp-high-energy}-\ref{sec:pertstabsub}.

\begin{figure}[tb]
  \centering
  \includegraphics[width=0.8\textwidth]{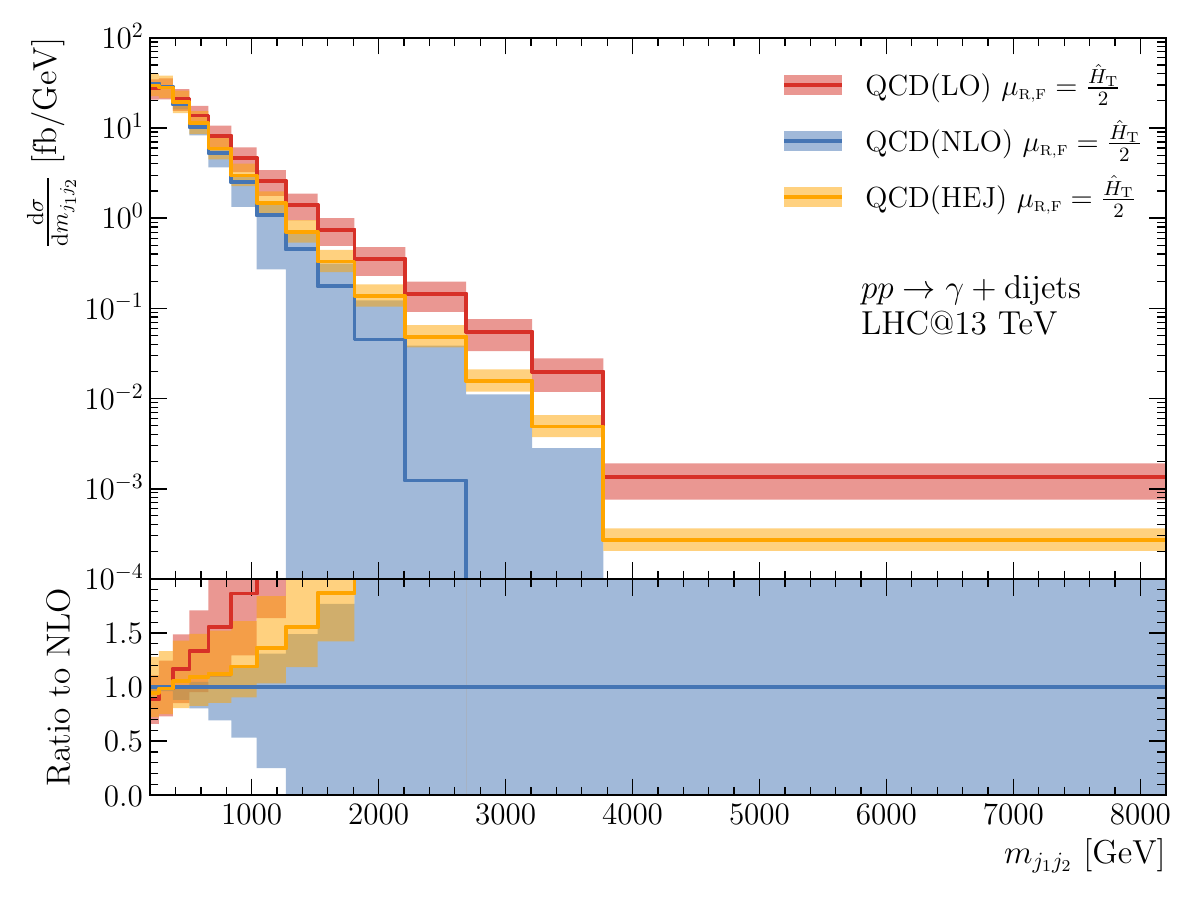}
  \caption{$\di\sigma/\di \mjj$ for pure QCD LO, NLO and HEJ matched to NLO}
  \label{fig:foHEJ}
\end{figure}
Figure~\ref{fig:foHEJ} shows in orange the result for $\di\sigma/\di\mjj$ obtained with the all-order summation and matching procedure for this process.
The result is in-between the fixed order curves for LO and NLO, and the scale variation is reduced compared to that found at LO.
The results all agree within scale variations at small \mjj, but diverge from around 1TeV.
Compared to the result at LO, the distribution falls of slightly steeper.
This fall-off and the shape of the distribution is directly controlled by the resummation.
In fact, the unphysical limiting behaviour of the NLO calculation can be understood from the NLO expansion of the resummation: having removed much of the real emission contribution with a central jet veto, the influence of the virtual corrections is seen immediately.
The high energy limit of the virtual corrections is dominated by a term $\propto\-\as \log(s/t)$, so it is hardly surprising that ultimately the cross section can become very small or even negative.
This divergent behaviour at NLO is regulated by the all-order resummation: The universal logarithmic virtual corrections are summed to an exponential and the real emission contributions calculated order by order with universal effective vertices; but the order-by-order expansion of the resummation still displays the same divergent behaviour seen at full fixed-order NLO, see appendix~\ref{sec:pertstabsub} for further details.

In conclusion, we have demonstrated that addressing the terms of order $\as\log(s)$ to all orders in the perturbation theory has regulated the divergent behaviour, which showed up at NLO.\footnote{To further identify the high-energy logarithmic terms as the cause of problems for the fixed-order expansion, we also calculate in appendix~\ref{sec:mjj} the fixed-order contributions using the factorisation and renormalisation scale \mjj (such that $\log(s/\mu_r^2)$ does not grow).
While the series appears more stable, the central scale choice leads to results which are at the edge of the scale variance band. This strongly suggests that scale variance is not a trustworthy indicator for missing higher order terms.
}
Obviously, the next question is how the resummed prediction compares then to data.

\section{Comparison to Data}
\label{sec:CompData}
Before comparing directly to data, the electro-weak component has to be added to the QCD component of the previous sections.
The electro-weak here was calculated with VBFNLO~3.0~\cite{Jager:2010aj,Baglio:2024gyp}.
The NLO QCD corrections to the EW process have a very different logarithmic structure from that of the QCD dominated process for similar reasons as those observed for Higgs production in association with dijets~\cite{Dokshitzer:1991he}.
We observe that compared to LO the QCD NLO correction to the electro-weak process is small at small \mjj, and from around 1TeV rises uniformly to 20\% in the bin of largest \mjj.

\begin{figure}[tb]
  \centering
  \includegraphics[width=0.8\textwidth]{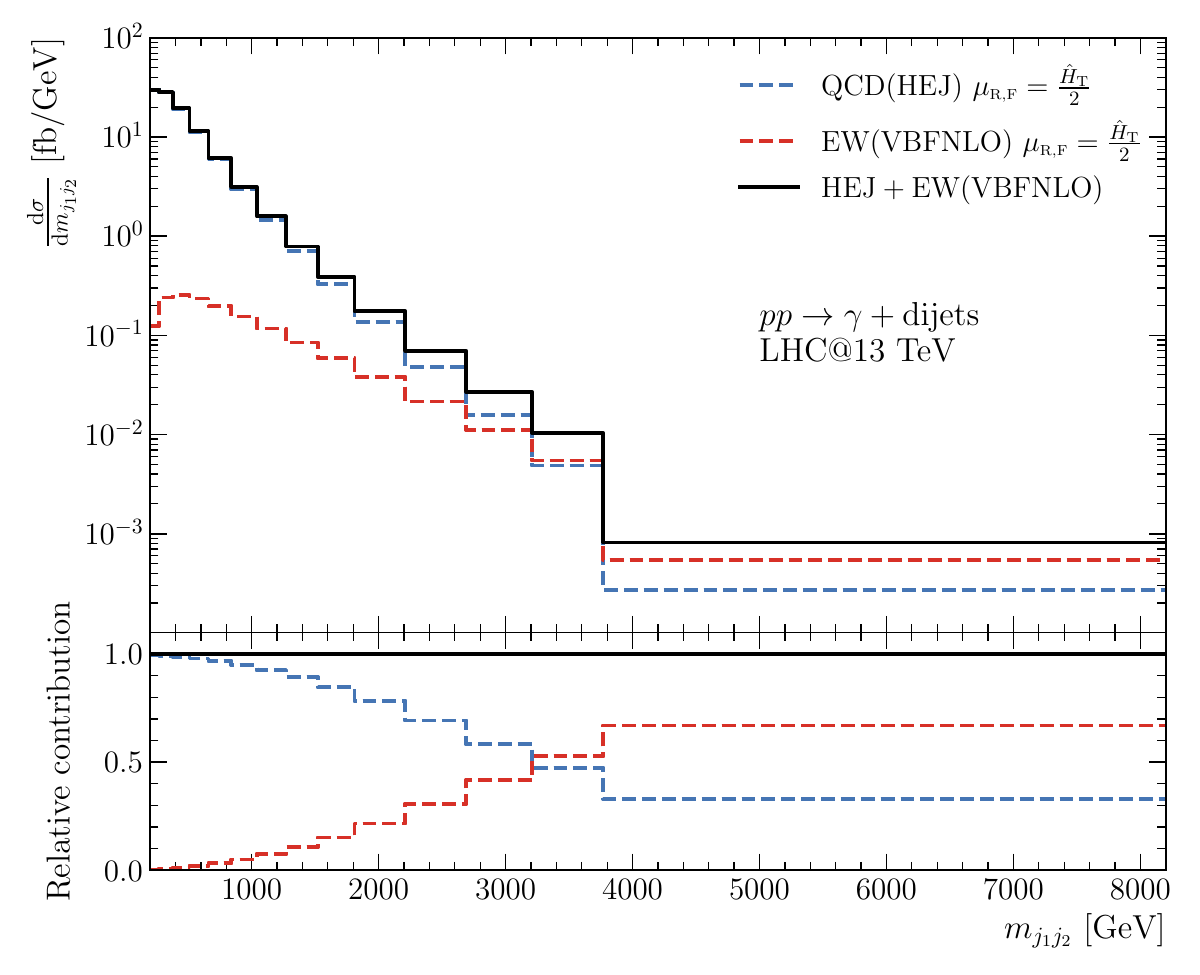}
  \caption{$\di\sigma/\di \mjj$ for the QCD component (calculated with HEJ) and the electro-weak component (calculated with VBFNLO). The process is dominated by the QCD contribution until around 3TeV.}
  \label{fig:thcomponents}
\end{figure}
Figure~\ref{fig:thcomponents} shows the QCD and the electroweak component of $pp\to\gamma jj$ as a function of \mjj.
The QCD component dominates for $\mjj\leq3$TeV dropping to 30\% of the cross section in the last bin.
The impact of any variation in the prediction of the QCD component will therefore decrease with increasing \mjj.

\begin{figure}[tb]
  \centering
  \begin{subfigure}{0.6\textwidth}
    \includegraphics[width=\textwidth,valign=t]{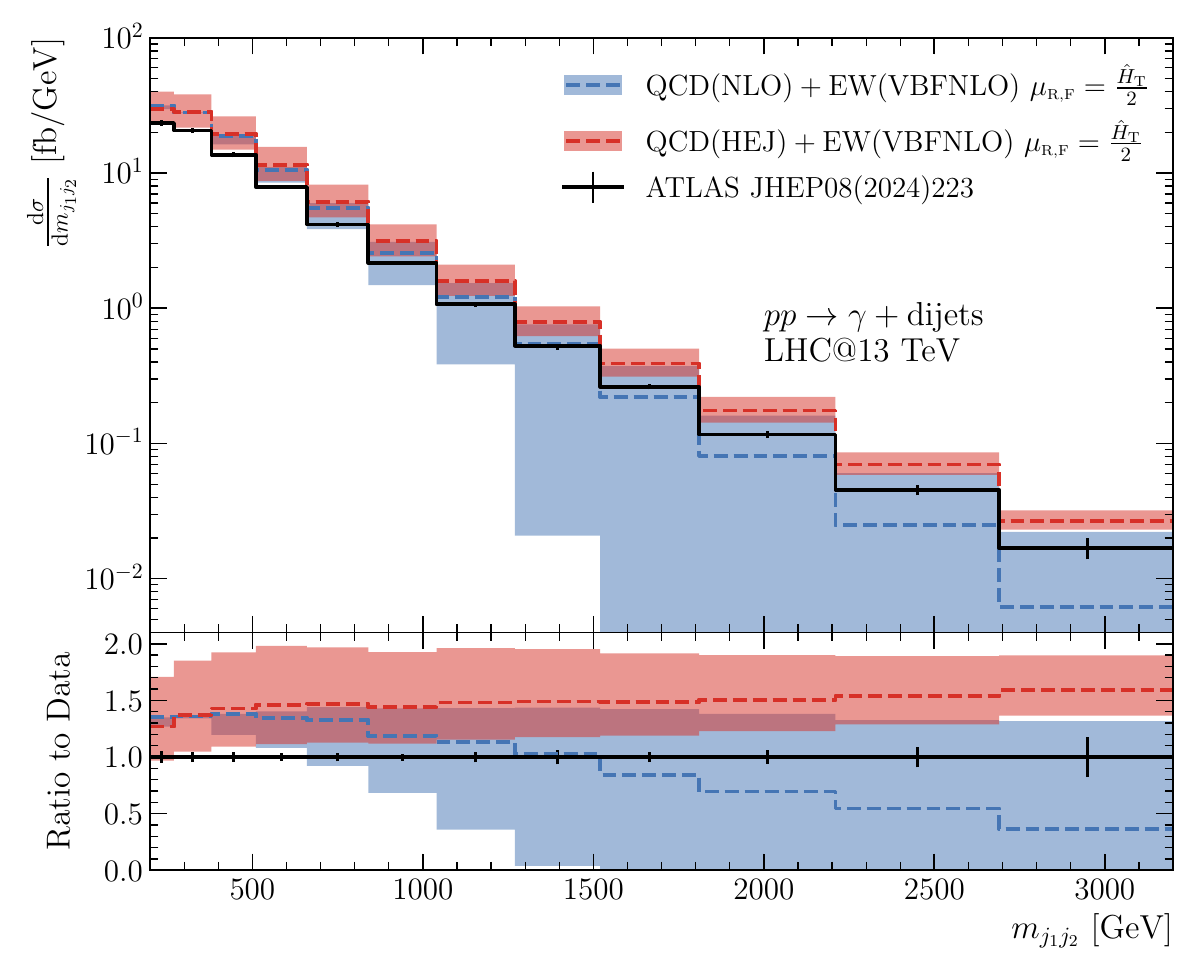}
  \end{subfigure}
  \begin{subfigure}{0.39\textwidth}
    \includegraphics[width=\textwidth,valign=t]{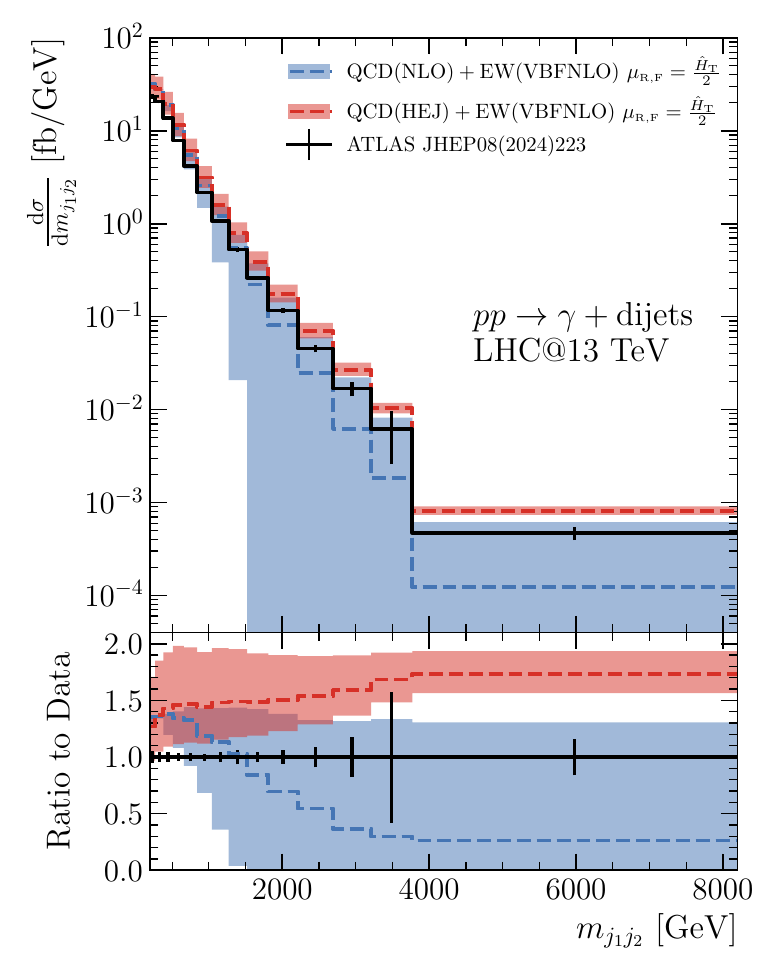}
  \end{subfigure}
  \caption{$\di\sigma/\di \mjj$ for NLO QCD+EW(VBFNLO), HEJ+EW(VBFNLO) and data. The prediction for HEJ is matched to NLO.}
  \label{fig:thvsdata5}
\end{figure}
With that in mind we compare in figure~\ref{fig:thvsdata5} data directly to the sum of the electro-weak component and the predictions of the QCD components calculated at NLO (blue) and with \HEJ (red).
Since the last two bins span half the range, and in order to emphasise the region dominated by the QCD process we present on figure~\ref{fig:thvsdata5} (left) the region $\mjj<3.2$TeV and on figure~\ref{fig:thvsdata5} (right) the full range.
We note that since the prediction from \HEJ inherits the normalisation from NLO it is overshooting the cross section by around 30\%.
The \HEJ prediction can easily be expanded to $\as^4\alpha$, so could have been matched to NNLO accuracy, had this calculation been performed for the current measurement.
Given the observation of the NNLO corrections~\cite{Badger:2023mgf} for the earlier more inclusive measurement~\cite{ATLAS:2019iaa} it is conceivable that the NNLO calculation would correct the normalisation.

The shape of the distribution with the combined prediction from \HEJ and VBFNLO is in very good agreement with data.
To illustrate this we investigate two methods for setting the normalisation of both the NLO and \HEJ predictions.
One can normalise the predictions to the data (equivalent to comparing $1/\sigma\ \di\sigma/\di\mjj$), or float the normalisation to minimise the $\chi^2$.
In table~\ref{tab:chi2} we list in the first column the scaling factors needed to normalise to data for QCD(NLO)+EW(NLO), QCD(HEJ)+EW(NLO) and the MEPS@NLO~\cite{Hoeche:2012yf} prediction included in the ATLAS study~\cite{ATLAS:2024vqf}.
The fact that the scaling factors for NLO(QCD+EW) and \HEJ+EW(NLO) are similar (0.755 and 0.716 respectively) is a result of matching \HEJ to the perturbative series at NLO.
In other words, the scaling factor for \HEJ is roughly the same with which NLO overshoots data.
The $\chi^2$-values obtained for the central predictions with this normalisation are 16.3 NLO(QCD+EW), 5.03 (MEPS@NLO) and 1.25 (HEJ+EW) respectively.\footnote{Only the experimental uncertainty is used for calculating the $\chi^2$, the scale variation around the central prediction is not used (it is not symmetric and certainly not Gaussian distributed).}

Next, we investigate the choice of normalising each prediction to minimise the $\chi^2$.
The scaling factors obtained with this method are listed in the third column of table~\ref{tab:chi2}.
The pair of scaling factors and $\chi^2$ values are (0.810,14.3) (NLO(QCD+EW)), (0.779,3.18) (MEPS@NLO) and (0.689,0.670) (HEJ+EW) respectively.
\begin{table}
  \centering
\begin{tabular}{l|cdcd}
Prediction & Scaling factor $\sigma$& \multicolumn{1}{c}{$\chi^2_\sigma/dof$} & Scaling factor $\chi^2_\textrm{min}$ & \multicolumn{1}{c}{$\chi^2_\textrm{min}/dof$} \\
\hline
QCD(NLO)+EW(NLO) & 0.755 & 16.3 & 0.810 & 14.3\\
MEPS@NLO & 0.832 & 5.03 & 0.779&3.18\\
QCD(HEJ)+EW(NLO) & 0.716 & 1.25 & 0.689 & 0.670\\
\hline
\end{tabular}
  \caption{Scaling factors for normalising the cross section and for minimising $\chi^2$ in the description of the distribution.}
  \label{tab:chi2}
\end{table}

\begin{figure}[tb]
  \centering
  \begin{subfigure}{0.6\textwidth}
    \includegraphics[width=\textwidth,valign=t]{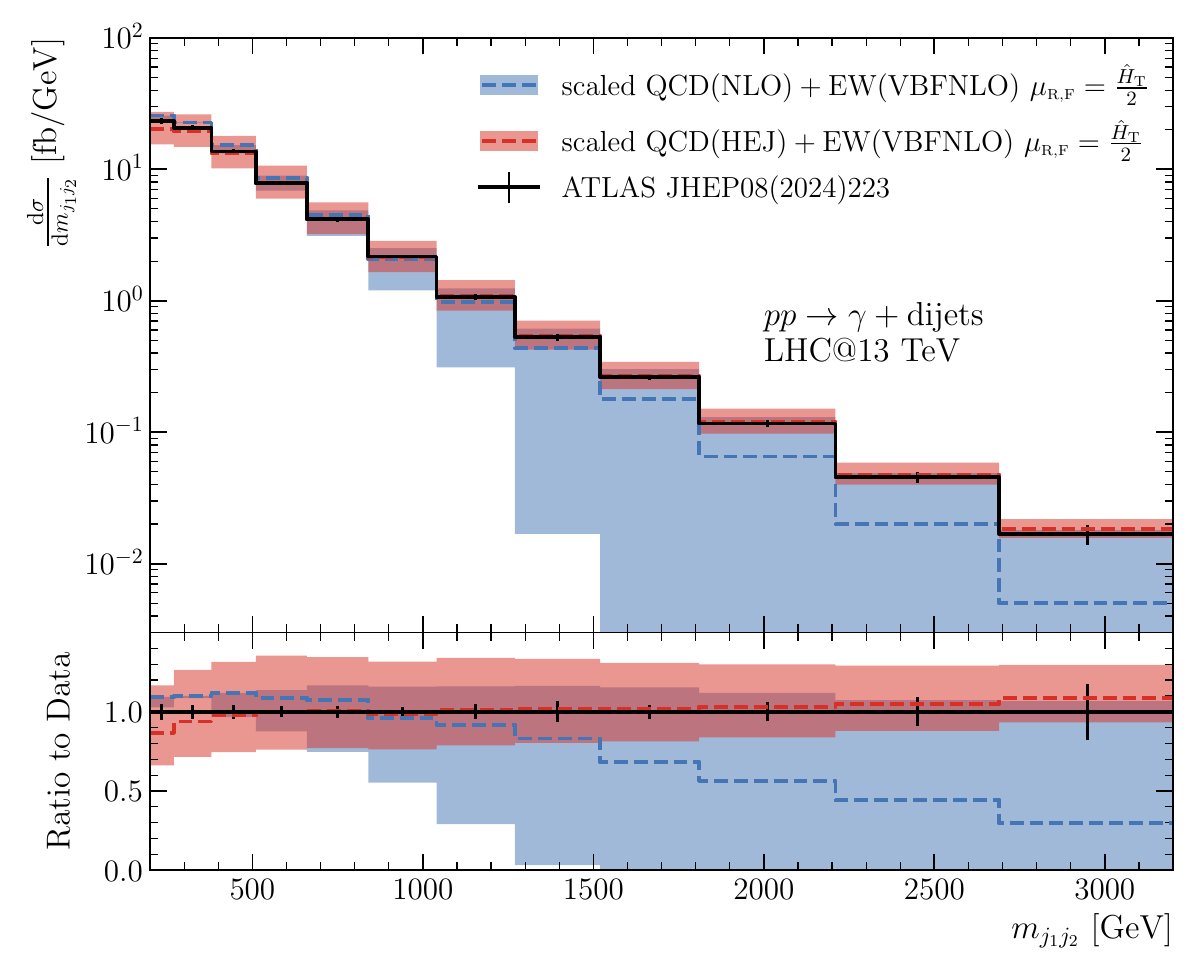}
  \end{subfigure}
  \begin{subfigure}{0.39\textwidth}
    \includegraphics[width=\textwidth,valign=t]{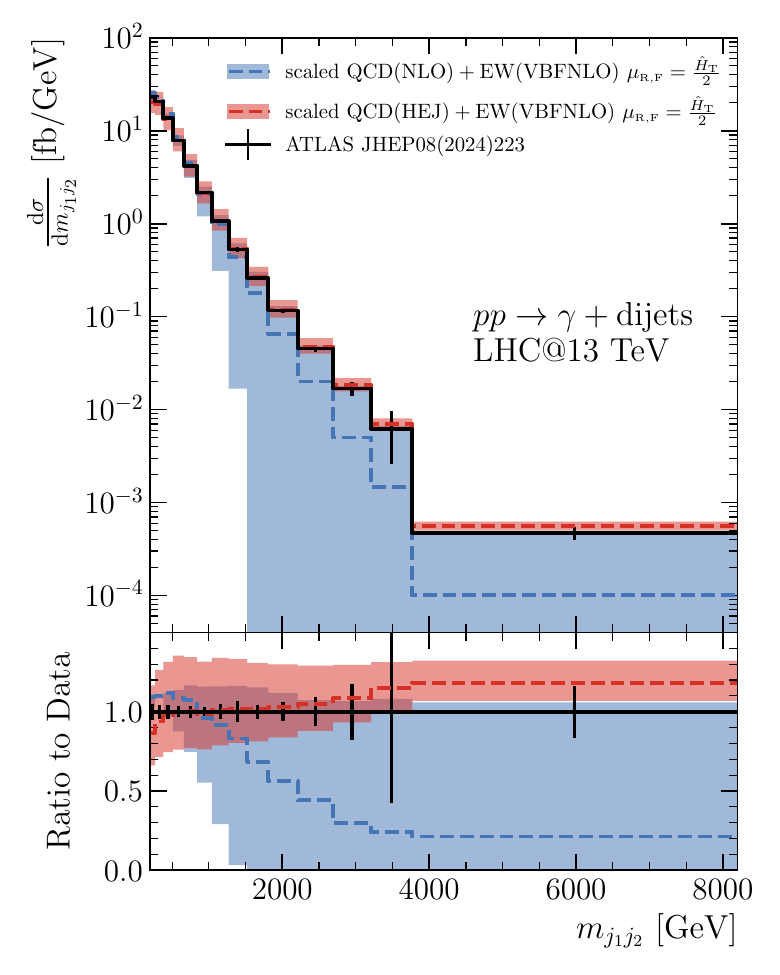}
  \end{subfigure}
  \caption{$\di\sigma/\di \mjj$ for scaled NLO, HEJ and data, see table~\ref{tab:chi2} for the scaling factors and $\chi^2/\mathrm{dof}$ obtained.}
  \label{fig:thvsdata3}
\end{figure}
The scaled predictions are presented in figure~\ref{fig:thvsdata3}.\footnote{Figure~\ref{fig:thvsdatamepsnlo} in appendix~\ref{sec:mepsnloresults} contains the figures with the result from reference~\cite{ATLAS:2024vqf} of MEPS@NLO included.
}
As is evident from there, the systematic treatment of the high-energy logarithms results in an excellent description of the distribution.
Judging from the findings in reference~\cite{Badger:2023mgf}, it is possible that the NNLO corrections on the NLO result would be a more accurate prediction of the normalisation, which could then be used for perturbative matching.

We note that the trends and conclusions are completely in line with those found in reference~\cite{ATLAS:2024vqf} for $Z$+dijets.
Compared to data, the normalisation for the predictions for $pp\to\gamma jj$ is a little higher than those for $pp\to Z jj$.
But the prediction for the shape of $\di\sigma/\di\mjj$ from the NLO matched and parton-shower merged MEPS@NLO fails to describe the shape for both $Z$ and $\gamma$ production in association with dijets.
The systematic treatment in \HEJ of the large logarithmic at increasing \mjj improves the description for all the cases studied.

\section{Conclusions}
\label{sec:Conclusions}

We identified issues in the fixed order predictions for a recent ATLAS measurement~\cite{ATLAS:2024vqf} as arising from high-energy logarithms.
These issues are solved in the systematic all-order logarithmic treatment within \HEJ.
The stabilised perturbative prediction give a good description of the distribution $\di\sigma/\di\mjj$ which was not well described by predictions available until now.
We note that these results are based on first-principle purely perturbative predictions without any tuning.

The analysis~\cite{ATLAS:2024vqf} thereby joins another recent measurement from ATLAS~\cite{ATLAS:2024png} indicating the appearance of high energy logarithmic effects in analyses relevant for standard perturbative observables and for BSM search and exclusions, in particular for the phase space regions normally  considered for studies of vector boson fusion and vector boson scattering.

The issues identified in this study are also present in the channels of $Z$+jets and $W$+jets considered in reference~\cite{ATLAS:2024vqf}.
These processes will be the focus of further studies.

The predictions presented in the current study were obtained using the public release of \HEJ~2.3 available at \href{http://hej.hepforge.org}{http://hej.hepforge.org}.

\acknowledgments

We would like to thank J.~Butterworth and C.~G\"utschow for their repeated encouragements to provide further predictions for the measurements in reference~\cite{ATLAS:2024vqf};
A.~Karlberg for discussions on the calculation of the electro-weak component and D.~Reichelt for advice on using Sherpa with MPI; and finally
L.~Dixon, N.~Glover, G.~Salam and members of CERN TH for discussions of some of the questions raised by these findings.
The authors express their thanks to current and previous collaborators of \HEJ.
The investigations reported here were made possible by the efficient management of the computing resources provided by the IPPP, University of Durham, and the UK GridPP~\cite{GridPP:2006wnd,Britton:2009ser} site \texttt{UKI-SCOTGRID-DURHAM}.
The work of Jeppe R.~Andersen is supported by the STFC under grant ST/X003167/1.
He would like to thank CERN~TH for kind hospitality for the duration of this work.
This work has been partially funded by the Eric \& Wendy Schmidt Fund for Strategic Innovation through the CERN Next Generation Triggers project under grant agreement number SIF-2023-004.
The work was also supported in part by the Spanish Ministry of Science and Innovation (PID2020-112965GB-I00,PID2023-146142NB-I00), and by the Departament de Recerca i Universities from Generalitat de Catalunya to the Grup de Recerca 00649 (Codi: 2021 SGR 00649). This project has received funding from the European Union’s Horizon 2020 research and innovation programme under grant agreement No 824093. IFAE is partially funded by the CERCA program of the Generalitat de Catalunya.

\appendix

\boldmath
\section{Components for the High Energy Resummation of $\gamma+$Dijets}
\label{sec:comp-high-energy}
\unboldmath
The calculation of the components needed for the high-energy corrections to the process $pp\to\gamma jj$ follows from the approaches outlined for $pp\to Z/\gamma^* jj$~\cite{Andersen:2016vkp} and $pp\to Wjj$~\cite{Andersen:2020yax} (in both cases a decay of the vector boson to charged fermions is implied).
The outline of the calculation will be described in this appendix.
Compared to the previous studies, the calculation presented here is simplified by the photon being on-shell, so instead of having a propagator and decay current the amplitude will have just a polarisation vector.
The advancements compared to the previous studies involving $\gamma^*/Z$+jets are that all the next-to-leading logarithmic real-emission components equivalent to those derived earlier for $pp\to Wjj$ are now applied in a process where quantum interference between a photon emission from each of the components is significant and needs to be taken into account in order to get a satisfactory description of the full amplitude~\cite{Andersen:2016vkp}.
This is particularly true at the very large \mjj considered in this study, where the di-quark perturbative initial state contributes significantly.
The interference of photon emission from any of quark lines in the process leads to interferences in the resummation over several different transverse momenta.
This is described in reference~\cite{Andersen:2016vkp}, but is further complicated by the additional NLL components taken into account leading to up to 3 quark lines with possible photon emission in the $pp\to \gamma+$4j component.
The full details of both this calculation for $pp\to \gamma jj$ and that for $pp\to (Z\to)\nu\nu$+dijets will appear in future publication.
The process of photon production in association with dijets is available in the version 2.3 release of \HEJ available at \href{http://hej.hepforge.org}{http://hej.hepforge.org}.

\subsection{Impact Currents and Central Emission}
\label{sec:impactcurrent}

The appearance of all-order logarithmic $\log(s/t)$ corrections in the perturbative description of $pp\to \gamma jj$ follows identical arguments to those for $pp\to jj$~\cite{Fadin:1975cb,Kuraev:1976ge,Kuraev:1977fs,Balitsky:1978ic,DelDuca:1995hf,DelDuca:1995zy,DelDuca:1999ha}.
Firstly, one observes that for fixed transverse momenta and $s_{cd},s_{d\gamma}\gg |s_{bd}|$ the scattering amplitude for $q_ag_b\to q_cg_d\gamma$ calculated at Born level factorises into two components depending on the momenta of $(p_a,p_c,p_\gamma)$ and $(p_b,p_d)$ only, contracted in colour and momentum space as if they exchanged a single gluon:
\begin{align}
  \includegraphics[valign=c,width=0.3\textwidth]{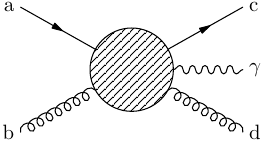}\ \overset{\longrightarrow}{\substack{s_{cd},s_{d\gamma}\gg |s_{bd}|\\\mathrm{fixed}\ p_t}}\ \includegraphics[valign=c,width=0.3\textwidth]{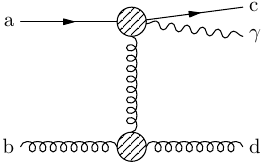}
\label{eq:amplitudesplit}
\end{align}
The requirement of fixed transverse momenta is to ensure the large invariant masses are reached by a hierarchy in positive and negative light-cone momenta, rather than by increasing transverse momenta.

Within the formalism of BFKL~\cite{Fadin:1975cb,Kuraev:1976ge,Kuraev:1977fs} the \emph{impact factor} of $(b,d)$ would depend on \emph{transverse} momenta only~\cite{DelDuca:1995zy,DelDuca:1999ha}.
In contrast, the factorised formalism developed for \HEJ~\cite{Andersen:2009nu,Andersen:2009he,Andersen:2011hs} maintains full momentum dependence, which ensures the analytic properties of the amplitude of crossing symmetry, gauge invariance and Lorentz invariance can be (and are) maintained.
Apart from the appeal of maintaining the analytic properties of the scattering amplitude, the matching corrections are smaller than for BFKL because of the much milder approximations.
Details of how the components are extracted from the full amplitude can be found in references~\cite{Andersen:2009nu,Andersen:2016vkp,Andersen:2020yax}.

The crucial conjecture~\cite{Fadin:1975cb,Kuraev:1976ge,Kuraev:1977fs,Balitsky:1978ic} for $pp\to jj$ is that the logarithmic $\log(s/t)$-component of the real part of the virtual corrections can be obtained to all orders by a simple substitution of the $t$-channel gluon propagator between the impact factors.
This conjecture has been verified directly up to next-to-leading logarithmic accuracy for full next-to-next-to-leading order amplitudes~\cite{DelDuca:2001gu}.
This means that the leading behaviour in the \emph{Regge kinematic limit} can be predicted to all orders.
In fact, not only processes dominated by the spin-1 colour octet exchange \emph{Reggeize} in this way, but so do processes with an exchange of quark quantum numbers~\cite{Bogdan:2002sr}.

The amplitudes for real emission corrections also simplify in the \emph{multi-Regge kinematic limit} (MRK) of increasing invariant mass between all of the final state coloured particles (again with fixed transverse momentum)~\cite{DelDuca:1999ha}.
In the MRK limit, each additional gluon emitted leads to a factor of the \emph{Lipatov vertex} $C_{\!A}\as/(\pi k_{i,\perp}^2)$ in the square of the amplitude.
This factorisation is illustrated in equation~(\ref{eq:amplitudesplit2}):
\begin{align}
  \includegraphics[valign=c,width=0.3\textwidth]{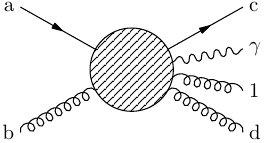}\ \overset{\longrightarrow}{\substack{s_{cd},s_{\gamma d}\gg s_{1d}\gg|s_{bd}|\\\mathrm{fixed}\ p_t}}\ \includegraphics[valign=c,width=0.3\textwidth]{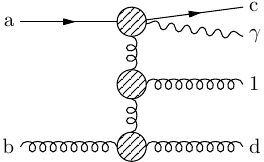}
\label{eq:amplitudesplit2}
\end{align}
In \HEJ, the central jet vertex is a Lorentz tensor~\cite{Andersen:2009nu} such that in the soft limit $k_{i,\perp}\to 0$ the effect on the square of the scattering amplitude tends to the simpler Lipatov vertex~\cite{Andersen:2011hs}.
However, the full tensor structure is of course necessary in order to fulfil the Ward identity.

For fixed $s_{cd}$, the phase space integral over gluon 1 leads to logarithmic corrections.
The infrared pole cancels between the contribution from the real emission and that of the virtual corrections, leading to a finite corrections which crucially has logarithmic contributions from both real and virtual corrections.
This feature persists in the MRK limit to all orders in the coupling.

\subsection{The Quasi Multi-Regge Kinematic Limit}
\label{sec:QMRK}
The factorisation of amplitudes into sub-components of less analytic complexity extends beyond leading logarithmic accuracy.
Specifically, factorised amplitudes can be constructed which describe correctly the situation when the restriction in equation~(\ref{eq:amplitudesplit2}) on the invariant masses is removed for one pair of coloured particles.
This situation depicted in equation~(\ref{eq:amplitudesplit3}) is termed the \emph{Quasi-Multi-Regge kinematic limit}.
\begin{align}
  \includegraphics[valign=c,width=0.3\textwidth]{figures/asymptote/phot3jetsamp.pdf}\ \overset{\longrightarrow}{\substack{s_{cd},s_{\gamma d},s_{1d}\gg|s_{bd}|\\\mathrm{fixed}\ p_t}}\ \includegraphics[valign=c,width=0.3\textwidth]{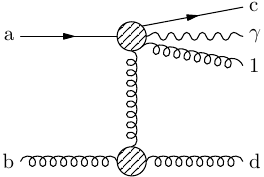}
\label{eq:amplitudesplit3}
\end{align}
Such components with leading contributions in the QMRK limit (and therefore NLL contributions in the MRK limit) are included in the current predictions.
We do not yet have the virtual corrections to the photon impact current in equation~(\ref{eq:amplitudesplit}), which is needed to regulate the IR gluon singularities in the impact current of equation~(\ref{eq:amplitudesplit3}).
Therefore, the impact current of equation~(\ref{eq:amplitudesplit3}) is used only when the quark and gluon are in separate jets, such that it belongs to the 3-jet (or more) contribution.

There are yet more complicated contributions from QMRK entering at four jets, in particular the emission of a photon from a central $q\bar q$ pair as illustrated in equation~(\ref{eq:amplitudesplit4}):
\begin{align}
  \includegraphics[valign=c,width=0.3\textwidth]{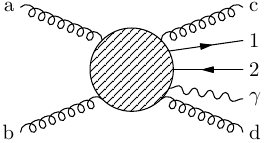}\ \overset{\longrightarrow}{\substack{s_{cd}\gg s_{ic},s_{id}\gg|s_{bd}|,|s_{ac}|\\i\in\{1,2,\gamma\},\ \mathrm{fixed}\ p_t}}\ \includegraphics[valign=c,width=0.3\textwidth]{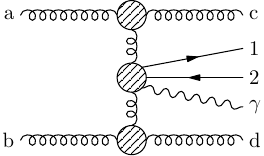}
\label{eq:amplitudesplit4}
\end{align}
The central vertex is extracted from full amplitudes as detailed in reference~\cite{Andersen:2020yax} for the case of $W$-emission.
A further complication in the present case of photon production is that for processes with more than one quark line, the photon can couple to all quark lines, and the interference between insertions on any line must be taken into account.
Each such interference generates separate momenta entering the resummation, since the $t$-channel momenta differ depending on whether the photon is emitted from an impact current or from a central emission vertex.
Such interference is fully taken into account in \HEJ as described in reference~\cite{Andersen:2016vkp} in the case of $Z$-production.

\subsection{Resummation}
\label{sec:resummation}
The simplicity of both the real and virtual corrections in the MRK and QMRK limit allow for all-order results to be obtained within this approximation.
The reliance on the asymptotic MRK limit of the amplitudes within BFKL theory, and the ensuing kinematic simplification to a dependence on 2-dimensional transverse momentum only, followed by a leading logarithmic accurate approximation to the phase space, allows for the fully inclusive all-order resummation of the emissions analytically through the BFKL~\cite{Balitsky:1978ic} equation.\footnote{The understanding of the factorisation of amplitudes is currently being pushed beyond even NLL~\cite{Buccioni:2024gzo,Abreu:2024xoh}.}

The milder approximations and added complexity of the Lorentz invariant amplitudes in \HEJ means that the phase space integrations have to be done numerically for each multiplicity, just as in the case of the full QCD amplitudes.
Each multiplicity is integrated over explicitly.
This has the obvious benefit over analytic resummation studies of allowing for the implementation of phase space cuts (e.g.~jet vetos) that are crucial for this study.
Technically, the all-order results are constructed using elements from the complexity of each of the references~\cite{Andersen:2011hs,Andersen:2016vkp,Andersen:2020yax}.

On top of the leading logarithms (denoted by \LL) and the leading logarithmic corrections to all the sub-leading channels (denoted \LLp) \HEJ includes for this calculation full matching of the Born amplitudes with up to 5 jets.
The higher jet rates normally associated with high-energy corrections are here suppressed because of the central jet veto.
In fact, the 3-jet contribution is less than 20\% of the cross section across all \mjj, and the 4 and 5 jet contributions are at the level of 5\% level and less than 1\% respectively.\footnote{The fact that the $\sigma_{3j}/\sigma_{2j}\approx\sigma_{4j}/\sigma_{3j}\approx\sigma_{5j}/\sigma_{4j}$ can be understood as the combined result of the central jet veto and the NLL contribution of unordered emissions, see appendix~\ref{sec:QMRK}.}
This ensures that for most of the cross section (namely for the two-jet rate), \mjj will be a good estimate for $\sqrt s$ especially for large \mjj.

\section{Perturbative Stability Including Subleading Effects}
\label{sec:pertstabsub}

Given the large variation in the perturbative spectrum between fixed leading order and fixed next-to-leading order, and of its own accord, it is interesting to study the perturbative stability in the spectrum predicted with \HEJ when taking into account the sub-leading channels described in appendix~\ref{sec:QMRK}.
The 3-jet component of these (as in equation~(\ref{eq:amplitudesplit3})) are channels that also enter at NLO, whereas the 4-jet component (as in equation~(\ref{eq:amplitudesplit4})) would enter at NNLO.
\begin{figure}[tb]
  \centering
  \includegraphics[width=0.8\textwidth]{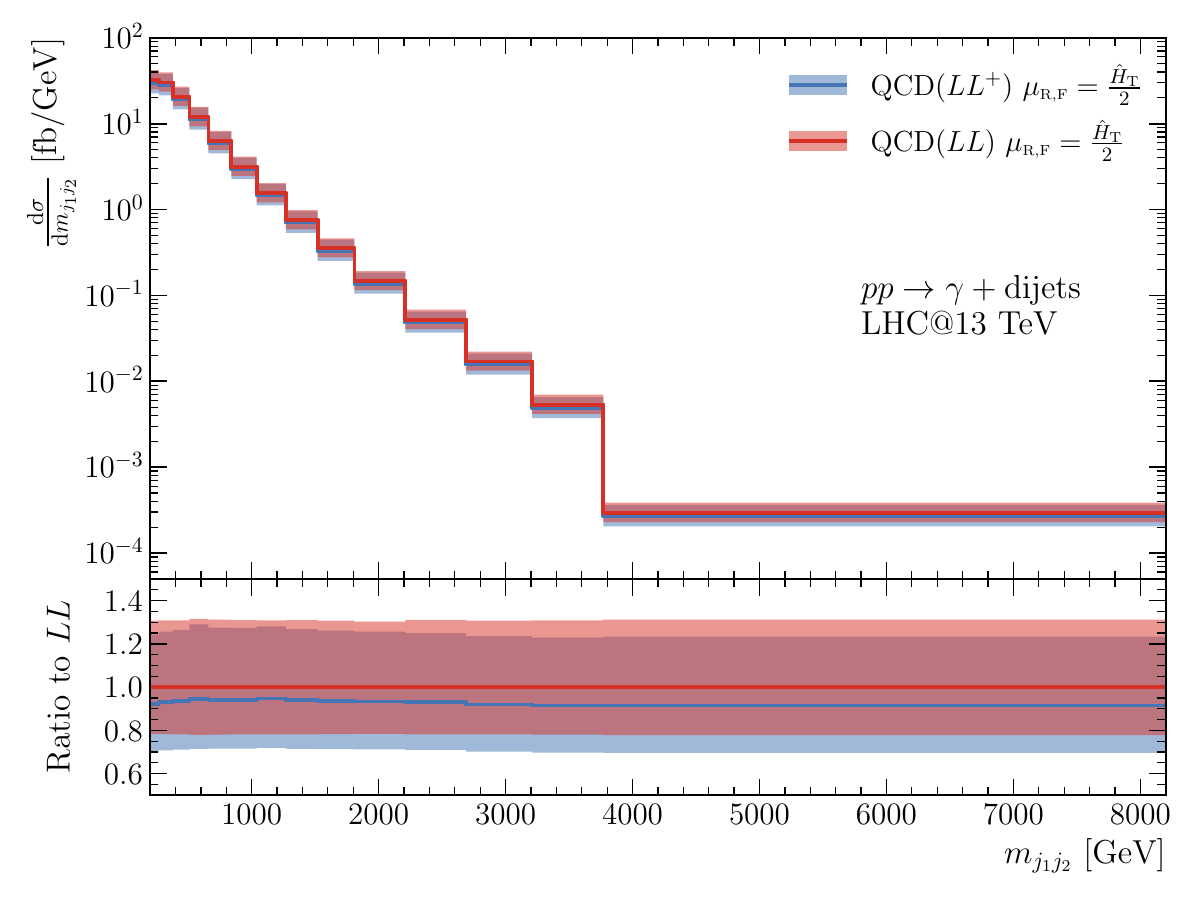}
  \caption{$\di\sigma/\di \mjj$ calculated with pure LL and including the sub-leading effects described in appendix~\ref{sec:QMRK}. The results were obtained using $\mu_R=\mu_F=\htt$ and the cross section is in each case matched to NLO.}
  \label{fig:follnllstable}
\end{figure}

Figure~\ref{fig:follnllstable} compares the QCD component of $pp\to\gamma$+dijets within the cuts of the current study (table~\ref{tab:cuts}) calculated with \HEJ at pure leading-logarithmic accuracy (\LL), and including also the sub-leading channels of the quasi-multi-Regge kinematic limit (\LLp).
The normalisation of both have been matched to NLO by calculating the cross section of the expansion of \LL and \LLp to the same accuracy $O(\alpha_s^3\alpha)$ as NLO, and multiplying the resummed distribution by the ratio of fixed order to the expanded result.
This procedure ensures NLO precision of the cross section and logarithmic accuracy of the distribution.
The differences between \LL and \LLp are less than 20\% and relatively uniform across \mjj.
We conclude that the distribution is perturbatively stable against the changes studied at sub-leading logarithmic order.
The systematic treatment of the $\log(s)$ corrections to all orders has stabilised the perturbative expansion.

\begin{figure}[tb]
  \centering
  \includegraphics[width=0.8\textwidth]{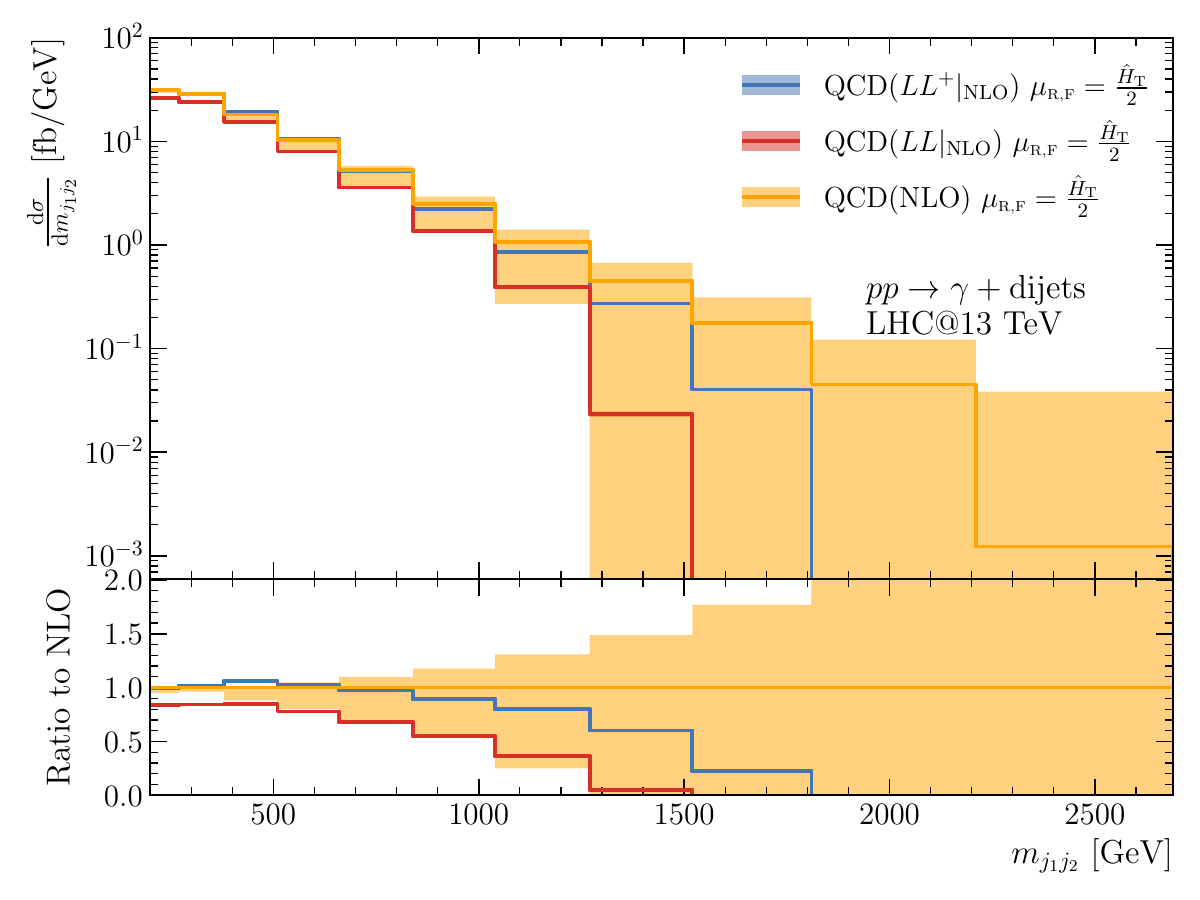}
  \caption{$\di\sigma/\di \mjj$ for full QCD NLO and \LL and \LLp expanded to $\alpha_s^3\alpha$.}
  \label{fig:llllpnlo}
\end{figure}

For completion, we compare in figure~\ref{fig:llllpnlo} $\di\sigma/\di\mjj$ for full QCD NLO, and for the expansions of \LL and \LLp to $\alpha_s^3\alpha$.
We note that \LLp is closer to the full NLO than \LL for all \mjj, and all three descriptions turn negative for large \mjj.
It may at first be surprising that the full NLO and \LL or \LLp terminated at order $\alpha_s^3\alpha$ do not converge at increasing \mjj.
This is caused by sub-leading effects - the resummation essentially controls the slope of the curve but not the offset in \mjj.

\section{Plots with MEPS@NLO}
\label{sec:mepsnloresults}
\begin{figure}[tb]
  \centering
  \begin{subfigure}{0.6\textwidth}
    \includegraphics[width=\textwidth,valign=t]{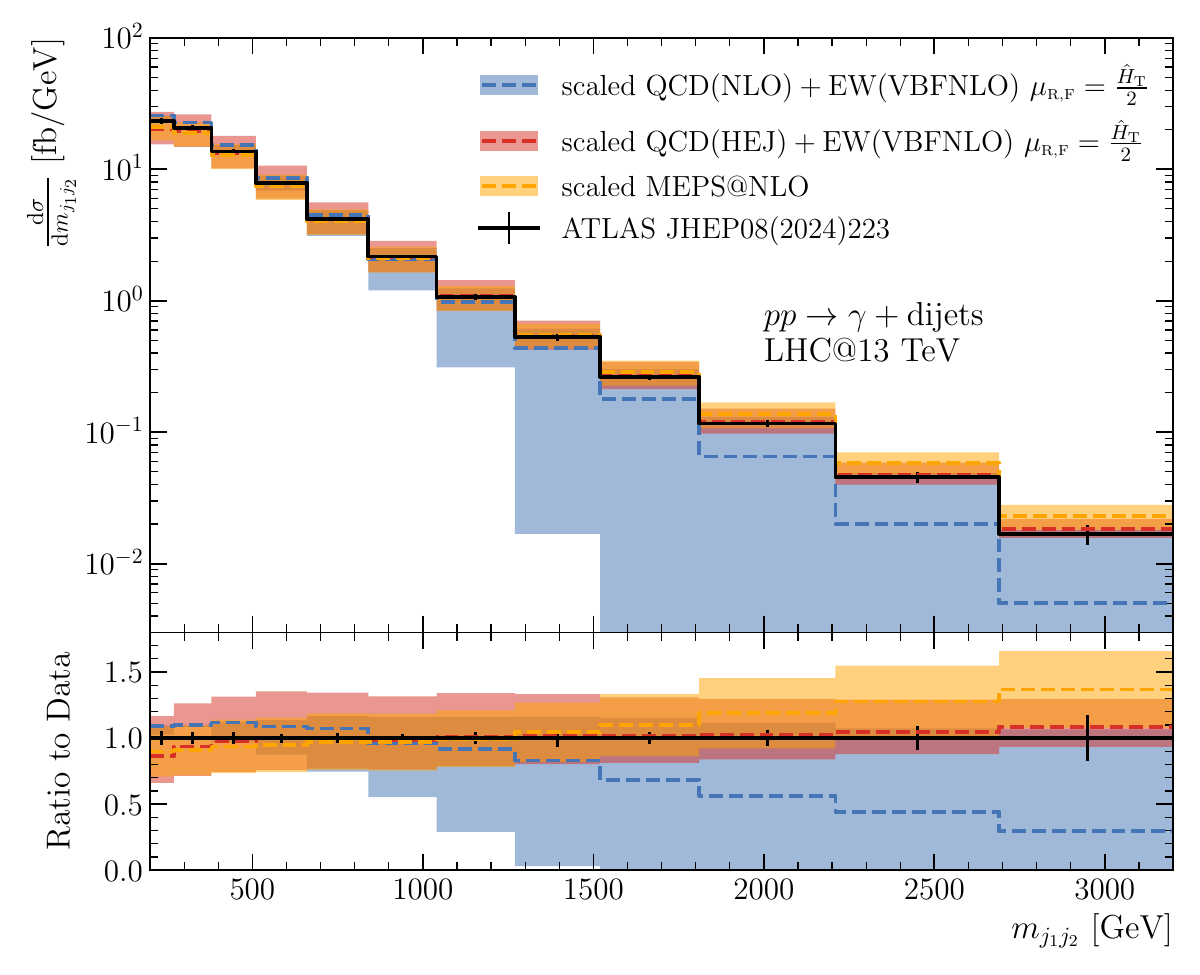}
  \end{subfigure}
  \begin{subfigure}{0.39\textwidth}
    \includegraphics[width=\textwidth,valign=t]{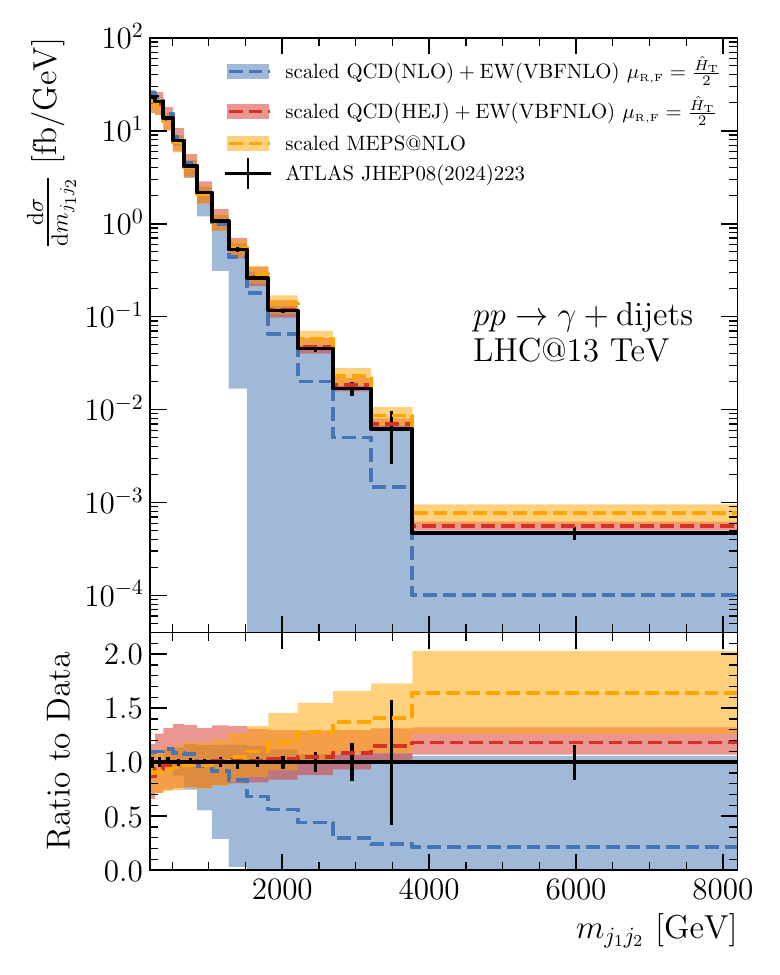}
  \end{subfigure}
  \caption{$\di\sigma/\di \mjj$ for scaled NLO, HEJ, MEPS@NLO and data, see table~\ref{tab:chi2} for the scaling factors and $\chi^2/\mathrm{dof}$ obtained}
  \label{fig:thvsdatamepsnlo}
\end{figure}

We present on figure~\ref{fig:thvsdatamepsnlo} the same results as in figure~\ref{fig:thvsdata3} but with the results of MEPS@NLO added. The results of MEPS@NLO were obtained from the HEPData~\cite{Maguire:2017ypu} entry of the ATLAS~\cite{ATLAS:2024vqf} study.

\boldmath
\section{Perturbative Results for Renormalisation and Factorisation Scale $\mu=\mjj$}
\label{sec:mjj}
\unboldmath

A natural choice for the factorisation and renormalisation scale even for large \mjj is based on the transverse momenta of jets and the photon.
Indeed, in the MRK limit, the $t$-channel factorisation discussed in appendix~\ref{sec:comp-high-energy} connects independent components of the amplitude with $t$-type momenta, where in the MRK limit $|t|\to p_t^2$.
Nevertheless, we have seen that the perturbative results obtained at NLO (or the NLO truncation of \HEJ) is perturbatively unstable with the scale choice \htt.

\begin{figure}[tb]
  \centering
  \includegraphics[width=0.8\textwidth]{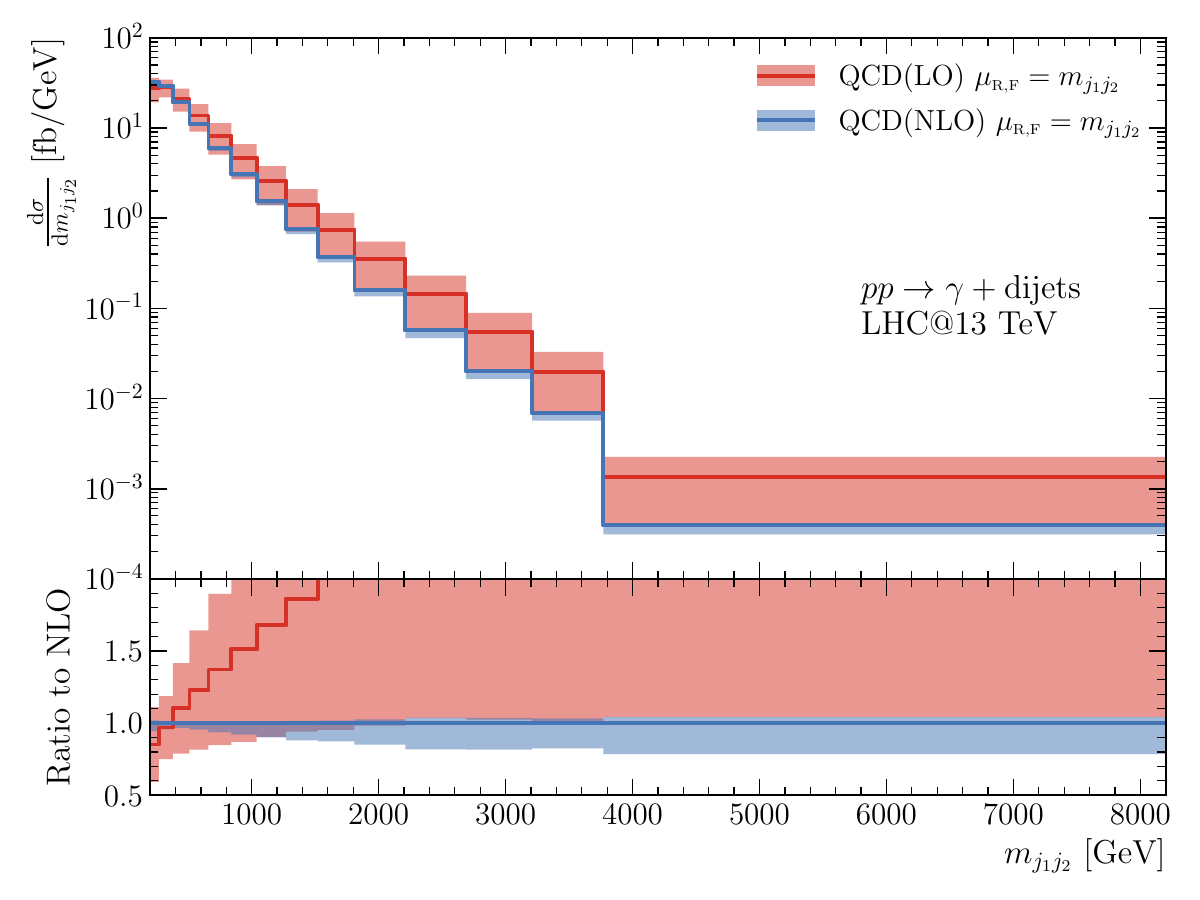}
  \caption{$\di\sigma/\di \mjj$ for pure QCD at LO and at NLO calculated with $\murf=\mjj$.}
  \label{fig:fomjj}
\end{figure}
\begin{figure}[tb]
  \centering
  \includegraphics[width=0.8\textwidth]{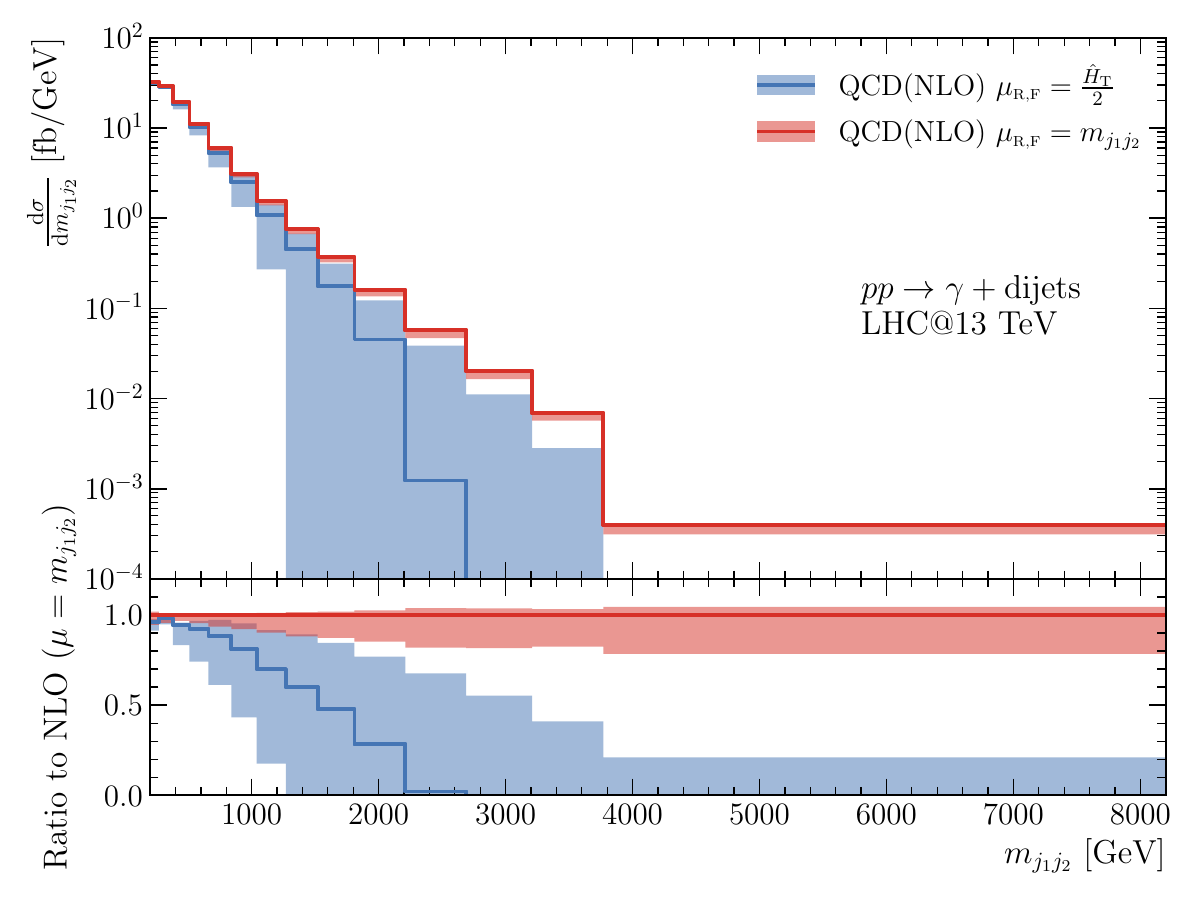}
  \caption{$\di\sigma/\di \mjj$ for QCD NLO with $\murf=\htt$ and $\murf=\mjj$. The scale variation bands for central scale choice of $\murf=\htt$ and $\murf=\mjj$ overlap only for small $\mjj$, and the central line for $\murf=\mjj$ is close to the extreme of the scale variation band around this central value. The central value for $\murf=\htt$ is negative for large $\mjj$.}
  \label{fig:nloht2vsmjj}
\end{figure}

We will here investigate another scale choice. Having identified the $\log(s)$ as a source of problems for the fixed-order perturbative description, it becomes natural to investigate the choice of $\murf=\mjj$.
Figure~\ref{fig:fomjj} shows the QCD fixed-order LO and NLO predictions (with scale variation bands obtained by the standard 7-point scale variation).
There appears to be better perturbative stability with this scale choice: The NLO central prediction is just at the edge of the LO scale variation band, and more importantly it remains positive for all \mjj.
The central NLO prediction is, however, close to the extreme of its scale variation band, which at least serves as a reminder that scale variation in itself is not necessarily a good proxy for estimating theory uncertainty.
Figure~\ref{fig:nloht2vsmjj} compares the NLO predictions for the two central scale choices of $\murf=\htt$ and $\murf=\mjj$.
The difference between these two scales grows for increasing \mjj and so do the predictions for the cross section, and there is hardly any overlap of the scale variation bands.
The interplay between effects from pdf and \as is interesting - even if the \as evaluation at the larger scale \mjj leads to a smaller \as, the evaluation of the pdfs at the larger scale means the cross section is uniformly larger for the scale choice $\murf=\mjj$.
In summary, choosing the scale $\murf=\mjj$ means that at least superficially the fixed-order perturbative predictions pass the standard test of perturbative stability, compared to the failure with $\murf=\htt$.

\begin{figure}[tb]
  \centering
  \includegraphics[width=0.8\textwidth]{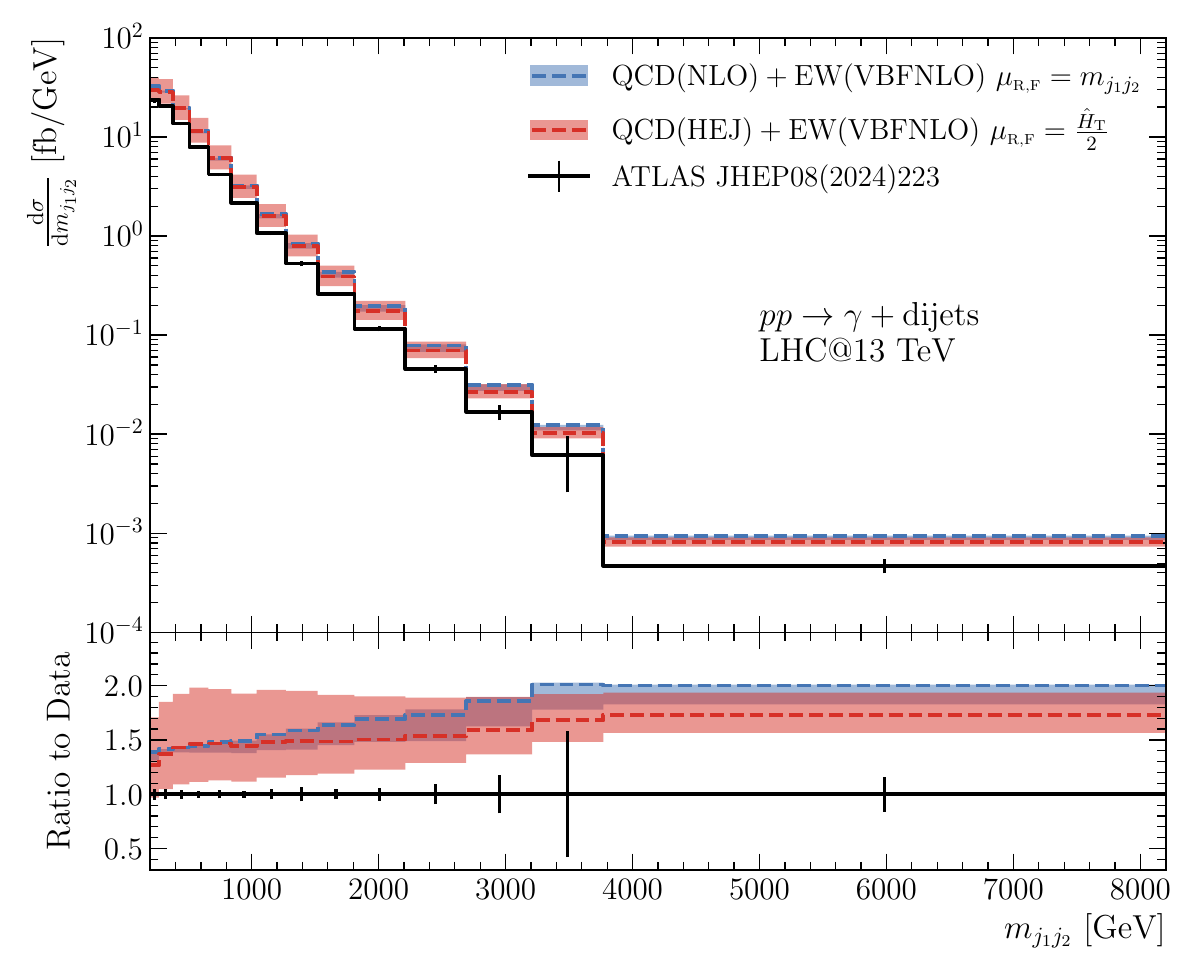}
  \caption{$\di\sigma/\di \mjj$ for NLO QCD+EWK with \mjj and HEJ+EWK (\htt) and data.}
  \label{fig:thvsdata2}
\end{figure}
The investigation of a scale choice of $\murf=\mjj$ for the fixed-order QCD component is pushed yet further in figure \ref{fig:thvsdata2} by adding the NLO VBF component to compare directly with data.
The figure includes also the same prediction from \HEJ presented on figure~\ref{fig:thvsdata5}.
The behaviour of the summed NLO predictions for the QCD($\murf=\mjj$) and EW($\murf=\htt$) components is surprisingly similar to that obtained with \HEJ, even if the spectrum is a little less steeply falling in \mjj.
We remind the reader that \HEJ systematically sums perturbative corrections of $\as\log(s)$ to all orders, which influence exactly the behaviour at large $\log(s)$ (and therefore at large $\log(\mjj)$).
Even if the shape of the NLO obtained with the larger scale choice is better than that obtained with $\murf=\htt$, the normalisation is still significantly above data.

\begin{figure}[tb]
  \centering
  \begin{subfigure}{0.6\textwidth}
    \includegraphics[width=\textwidth,valign=t]{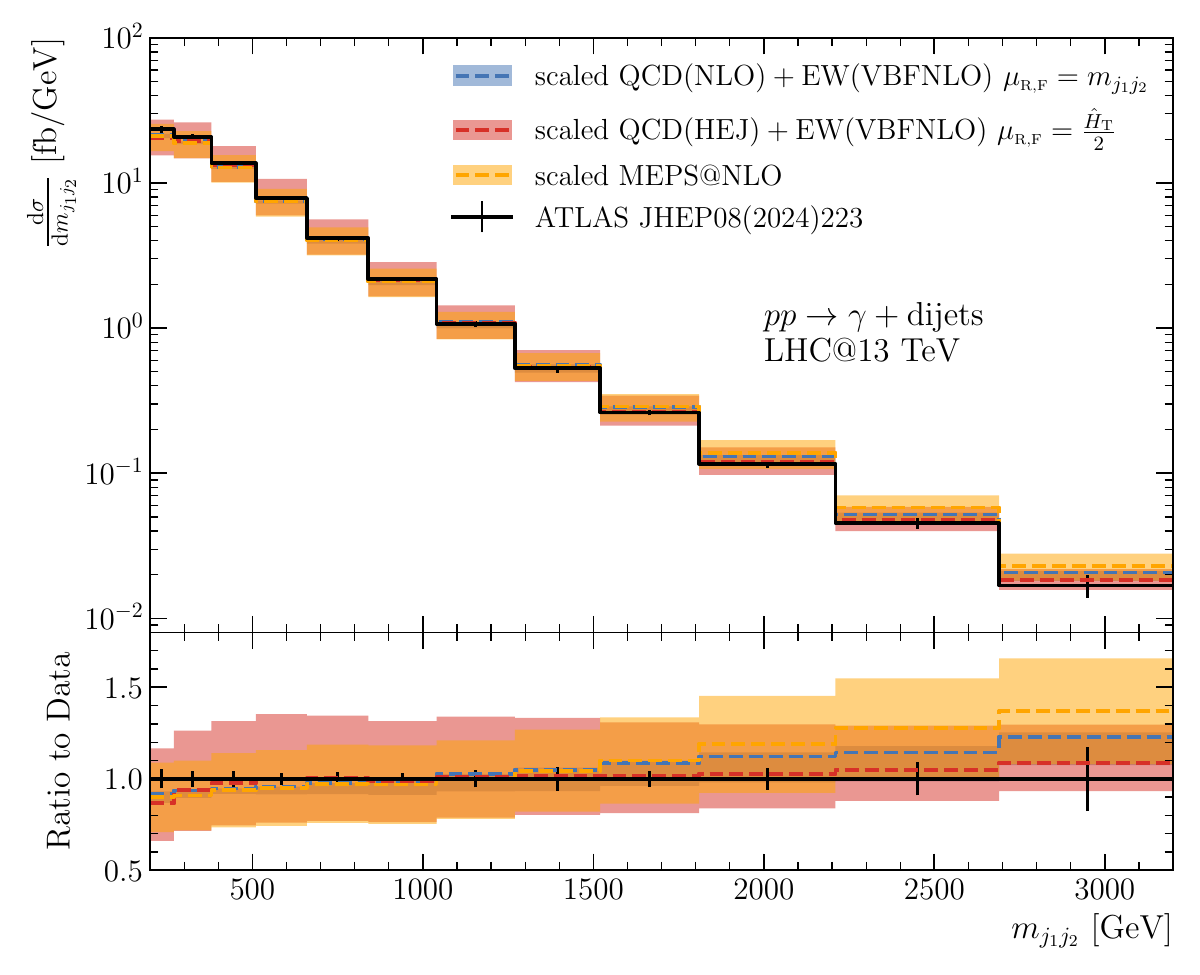}
  \end{subfigure}
  \begin{subfigure}{0.39\textwidth}
    \includegraphics[width=\textwidth,valign=t]{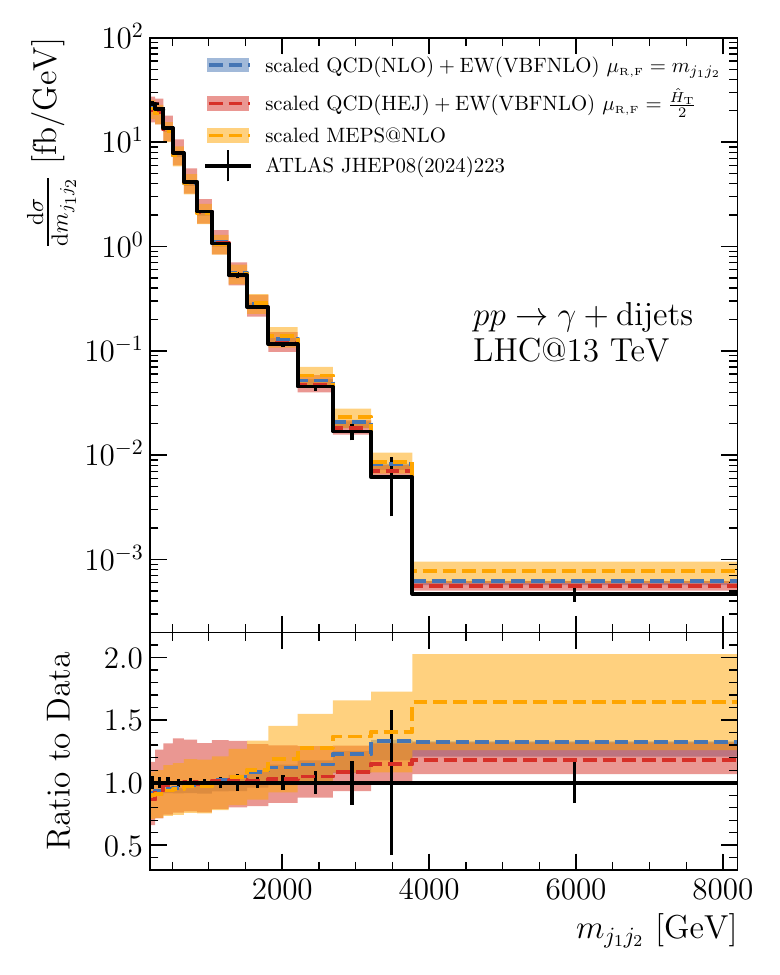}
  \end{subfigure}
  \caption{$\di\sigma/\di \mjj$ for scaled NLO $\murf=\mjj$, HEJ, MEPS@NLO and data.}
  \label{fig:thvsdata4}
\end{figure}
Figure~\ref{fig:thvsdata4} compares the NLO prediction normalised to minimise the $\chi^2$ with the scaling factor listed in table~\ref{tab:chi2b}.
Also shown are the predictions from MEPS@NLO (from the ATLAS paper~\cite{ATLAS:2024vqf}) and \HEJ already presented in figure~\ref{fig:thvsdata3}.

\begin{table}
  \centering
\begin{tabular}{l|cdcd}
Prediction & Scaling factor $\sigma$& \multicolumn{1}{c}{$\chi^2_\sigma/dof$} & Scaling factor $\chi^2_\textrm{min}$ & \multicolumn{1}{c}{$\chi^2_\textrm{min}/dof$} \\
\hline
QCD(NLO)+EW(NLO) & 0.755 & 16.3 & 0.810 & 14.3\\
MEPS@NLO & 0.832 & 5.03 & 0.779&3.18\\
QCD(NLO, $\murf=\mjj$)+EW(NLO) & 0.694 & 2.50 & 0.662  & 1.54\\
QCD(HEJ)+EW(NLO) & 0.716 & 1.25 & 0.689 & 0.670\\
\hline
\end{tabular}
  \caption{Scaling factors for normalising the cross section and for minimising $\chi^2$ in the description of the distribution.}
  \label{tab:chi2b}
\end{table}
The minimum $\chi^2/dof$ obtained is just twice that obtained with \HEJ and reduced by an order of magnitude compared to the result obtained with $\murf=\htt$.

\begin{figure}[tb]
  \centering
  \includegraphics[width=0.8\textwidth]{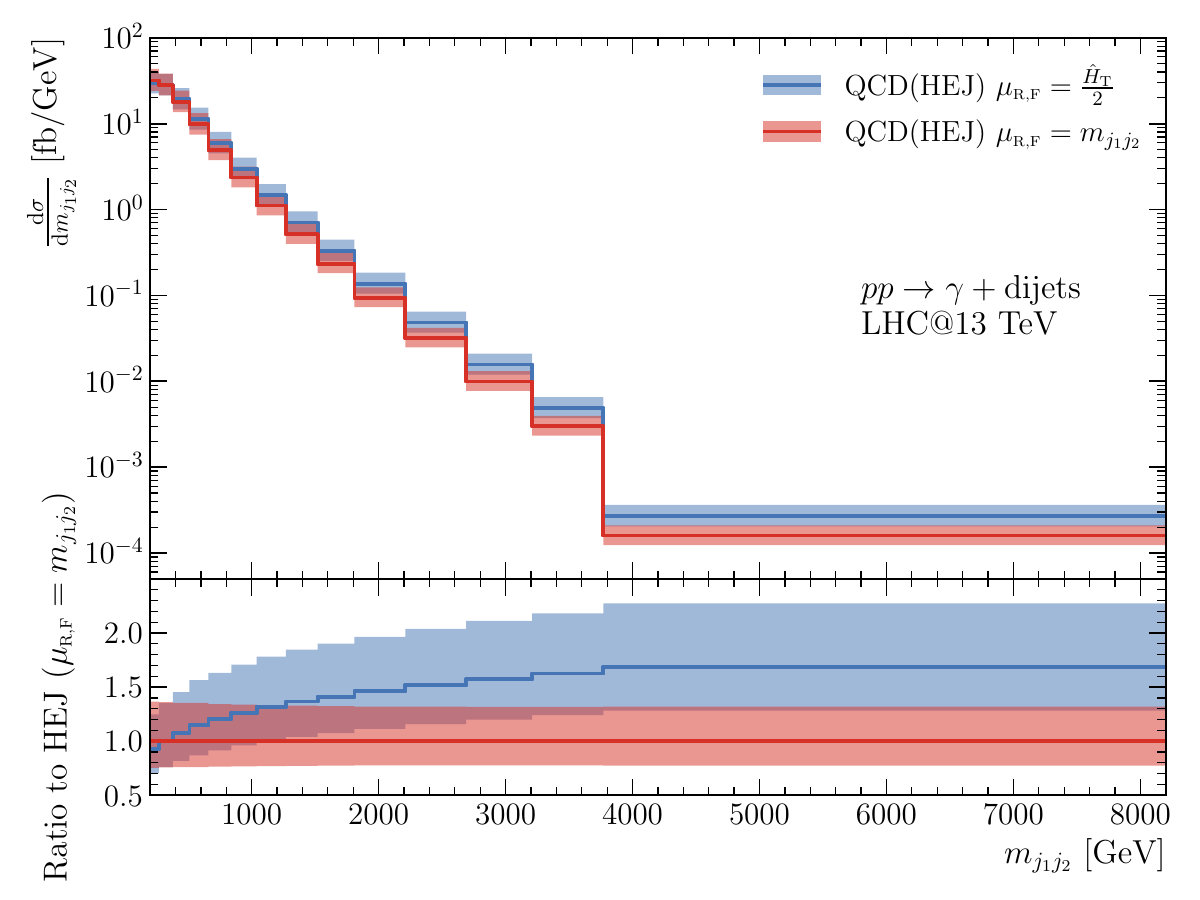}
  \caption{$\di\sigma/\di \mjj$ for pure QCD HEJ with $\murf=\htt$ and $\murf=\mjj$.}
  \label{fig:HEJht2vsmjj}
\end{figure}
Finally, to further indicate that the resummation in \HEJ of the $\log(s)$ perturbative corrections stabilises the perturbative prediction, figure~\ref{fig:HEJht2vsmjj} shows the resummed predictions obtained with the two scales choices $\murf=\htt$ and $\murf=\mjj$.
We emphasise again that especially in the approach with the amplitudes factorised into high-energy components, there is no physical argument for a renormalisation scale of $\murf=\mjj$.
Nevertheless, it is interesting that we observe an overlap between the scale variation bands.
Unsurprisingly, the difference between the predictions increases with increasing \mjj as the scale of \as is influenced.

\bibliographystyle{JHEP}
\bibliography{main.bib}

\providecommand{\href}[2]{#2}\begingroup\raggedright\begin{thebibliography}{10}

\bibitem{ATLAS:2024cgh}
{\scshape ATLAS} collaboration, \emph{{Standard Model Summary Plots June
  2024}}, {\emph{ATL-PHYS-PUB-2024-011} (2024) }.

\bibitem{ATLAS:2024vqf}
{\scshape ATLAS} collaboration, \emph{{Differential cross-sections for events
  with missing transverse momentum and jets measured with the ATLAS detector in
  13 TeV proton-proton collisions}},
  \href{https://doi.org/10.1007/JHEP08(2024)223}{\emph{JHEP} {\bfseries 08}
  (2024) 223} [\href{https://arxiv.org/abs/2403.02793}{{\ttfamily
  2403.02793}}].

\bibitem{ATLAS:2019iaa}
{\scshape ATLAS} collaboration, \emph{{Measurement of isolated-photon plus
  two-jet production in $pp$ collisions at $\sqrt s=13$ TeV with the ATLAS
  detector}}, \href{https://doi.org/10.1007/JHEP03(2020)179}{\emph{JHEP}
  {\bfseries 03} (2020) 179}
  [\href{https://arxiv.org/abs/1912.09866}{{\ttfamily 1912.09866}}].

\bibitem{Badger:2023mgf}
S.~Badger, M.~Czakon, H.B.~Hartanto, R.~Moodie, T.~Peraro, R.~Poncelet et~al.,
  \emph{{Isolated photon production in association with a jet pair through
  next-to-next-to-leading order in QCD}},
  \href{https://doi.org/10.1007/JHEP10(2023)071}{\emph{JHEP} {\bfseries 10}
  (2023) 071} [\href{https://arxiv.org/abs/2304.06682}{{\ttfamily
  2304.06682}}].

\bibitem{Hoeche:2012yf}
S.~Hoeche, F.~Krauss, M.~Schönherr and F.~Siegert, \emph{{QCD matrix elements
  + parton showers: The NLO case}},
  \href{https://doi.org/10.1007/JHEP04(2013)027}{\emph{JHEP} {\bfseries 04}
  (2013) 027} [\href{https://arxiv.org/abs/1207.5030}{{\ttfamily 1207.5030}}].

\bibitem{Catani:2013oma}
S.~Catani, M.~Fontannaz, J.P.~Guillet and E.~Pilon, \emph{{Isolating Prompt
  Photons with Narrow Cones}},
  \href{https://doi.org/10.1007/JHEP09(2013)007}{\emph{JHEP} {\bfseries 09}
  (2013) 007} [\href{https://arxiv.org/abs/1306.6498}{{\ttfamily 1306.6498}}].

\bibitem{Balsiger:2018ezi}
M.~Balsiger, T.~Becher and D.Y.~Shao, \emph{{Non-global logarithms in jet and
  isolation cone cross sections}},
  \href{https://doi.org/10.1007/JHEP08(2018)104}{\emph{JHEP} {\bfseries 08}
  (2018) 104} [\href{https://arxiv.org/abs/1803.07045}{{\ttfamily
  1803.07045}}].

\bibitem{Becher:2022rhu}
T.~Becher, S.~Favrod and X.~Xu, \emph{{QCD anatomy of photon isolation}},
  \href{https://doi.org/10.1007/JHEP01(2023)005}{\emph{JHEP} {\bfseries 01}
  (2023) 005} [\href{https://arxiv.org/abs/2208.01554}{{\ttfamily
  2208.01554}}].

\bibitem{Becher:2023mtx}
T.~Becher, M.~Neubert, D.Y.~Shao and M.~Stillger, \emph{{Factorization of
  non-global LHC observables and resummation of super-leading logarithms}},
  \href{https://doi.org/10.1007/JHEP12(2023)116}{\emph{JHEP} {\bfseries 12}
  (2023) 116} [\href{https://arxiv.org/abs/2307.06359}{{\ttfamily
  2307.06359}}].

\bibitem{Boer:2024xzy}
P.~B\"oer, P.~Hager, M.~Neubert, M.~Stillger and X.~Xu, \emph{{Resummation of
  Glauber phases in non-global LHC observables for large N$_{c}$}},
  \href{https://doi.org/10.1007/JHEP08(2024)036}{\emph{JHEP} {\bfseries 08}
  (2024) 036} [\href{https://arxiv.org/abs/2407.01691}{{\ttfamily
  2407.01691}}].

\bibitem{Becher:2024nqc}
T.~Becher, P.~Hager, G.~Martinelli, M.~Neubert, D.~Schwienbacher and
  M.~Stillger, \emph{{Super-leading logarithms in pp \textrightarrow{} 2
  jets}}, \href{https://doi.org/10.1007/JHEP01(2025)171}{\emph{JHEP} {\bfseries
  01} (2025) 171} [\href{https://arxiv.org/abs/2411.12742}{{\ttfamily
  2411.12742}}].

\bibitem{Boer:2024hzh}
P.~B\"oer, P.~Hager, M.~Neubert, M.~Stillger and X.~Xu,
  \emph{{Renormalization-group improved resummation of super-leading
  logarithms}}, \href{https://doi.org/10.1007/JHEP08(2024)035}{\emph{JHEP}
  {\bfseries 08} (2024) 035}
  [\href{https://arxiv.org/abs/2405.05305}{{\ttfamily 2405.05305}}].

\bibitem{Forshaw:2006fk}
J.R.~Forshaw, A.~Kyrieleis and M.H.~Seymour, \emph{{Super-leading logarithms in
  non-global observables in QCD}},
  \href{https://doi.org/10.1088/1126-6708/2006/08/059}{\emph{JHEP} {\bfseries
  08} (2006) 059} [\href{https://arxiv.org/abs/hep-ph/0604094}{{\ttfamily
  hep-ph/0604094}}].

\bibitem{Forshaw:2008cq}
J.R.~Forshaw, A.~Kyrieleis and M.H.~Seymour, \emph{{Super-leading logarithms in
  non-global observables in QCD: Colour basis independent calculation}},
  \href{https://doi.org/10.1088/1126-6708/2008/09/128}{\emph{JHEP} {\bfseries
  09} (2008) 128} [\href{https://arxiv.org/abs/0808.1269}{{\ttfamily
  0808.1269}}].

\bibitem{AngelesMartinez:2018cfz}
R.~\'Angeles~Mart\'\i{}nez, M.~De~Angelis, J.R.~Forshaw, S.~Pl\"atzer and
  M.H.~Seymour, \emph{{Soft gluon evolution and non-global logarithms}},
  \href{https://doi.org/10.1007/JHEP05(2018)044}{\emph{JHEP} {\bfseries 05}
  (2018) 044} [\href{https://arxiv.org/abs/1802.08531}{{\ttfamily
  1802.08531}}].

\bibitem{Banfi:2021xzn}
A.~Banfi, F.A.~Dreyer and P.F.~Monni, \emph{{Higher-order non-global logarithms
  from jet calculus}},
  \href{https://doi.org/10.1007/JHEP03(2022)135}{\emph{JHEP} {\bfseries 03}
  (2022) 135} [\href{https://arxiv.org/abs/2111.02413}{{\ttfamily
  2111.02413}}].

\bibitem{Becher:2024kmk}
T.~Becher, P.~Hager, S.~Jaskiewicz, M.~Neubert and D.~Schwienbacher,
  \emph{{Factorization Restoration through Glauber Gluons}},
  \href{https://doi.org/10.1103/PhysRevLett.134.061901}{\emph{Phys. Rev. Lett.}
  {\bfseries 134} (2025) 061901}
  [\href{https://arxiv.org/abs/2408.10308}{{\ttfamily 2408.10308}}].

\bibitem{Bierlich:2024vqo}
C.~Bierlich, A.~Buckley, J.M.~Butterworth, C.~Gutschow, L.~Lonnblad, T.~Procter
  et~al., \emph{{Robust independent validation of experiment and theory: Rivet
  version 4 release note}},
  \href{https://doi.org/10.21468/SciPostPhysCodeb.36}{\emph{SciPost Phys.
  Codeb.} {\bfseries 36} (2024) 1}
  [\href{https://arxiv.org/abs/2404.15984}{{\ttfamily 2404.15984}}].

\bibitem{Andersen:2007mp}
J.R.~Andersen, T.~Binoth, G.~Heinrich and J.M.~Smillie, \emph{{Loop induced
  interference effects in Higgs Boson plus two jet production at the LHC}},
  \href{https://doi.org/10.1088/1126-6708/2008/02/057}{\emph{JHEP} {\bfseries
  02} (2008) 057} [\href{https://arxiv.org/abs/0709.3513}{{\ttfamily
  0709.3513}}].

\bibitem{Bredenstein:2008tm}
A.~Bredenstein, K.~Hagiwara and B.~Jäger, \emph{{Mixed QCD-electroweak
  contributions to Higgs-plus-dijet production at the LHC}},
  \href{https://doi.org/10.1103/PhysRevD.77.073004}{\emph{Phys. Rev. D}
  {\bfseries 77} (2008) 073004}
  [\href{https://arxiv.org/abs/0801.4231}{{\ttfamily 0801.4231}}].

\bibitem{Dixon:2009uk}
L.J.~Dixon and Y.~Sofianatos, \emph{{Analytic one-loop amplitudes for a Higgs
  boson plus four partons}},
  \href{https://doi.org/10.1088/1126-6708/2009/08/058}{\emph{JHEP} {\bfseries
  08} (2009) 058} [\href{https://arxiv.org/abs/0906.0008}{{\ttfamily
  0906.0008}}].

\bibitem{Sherpa:2019gpd}
{\scshape Sherpa} collaboration, \emph{{Event Generation with Sherpa 2.2}},
  \href{https://doi.org/10.21468/SciPostPhys.7.3.034}{\emph{SciPost Phys.}
  {\bfseries 7} (2019) 034} [\href{https://arxiv.org/abs/1905.09127}{{\ttfamily
  1905.09127}}].

\bibitem{Buccioni:2019sur}
F.~Buccioni, J.-N.~Lang, J.M.~Lindert, P.~Maierh\"ofer, S.~Pozzorini, H.~Zhang
  et~al., \emph{{OpenLoops 2}},
  \href{https://doi.org/10.1140/epjc/s10052-019-7306-2}{\emph{Eur. Phys. J. C}
  {\bfseries 79} (2019) 866}
  [\href{https://arxiv.org/abs/1907.13071}{{\ttfamily 1907.13071}}].

\bibitem{Cacciari:2008gp}
M.~Cacciari, G.P.~Salam and G.~Soyez, \emph{{The anti-$k_t$ jet clustering
  algorithm}}, \href{https://doi.org/10.1088/1126-6708/2008/04/063}{\emph{JHEP}
  {\bfseries 04} (2008) 063} [\href{https://arxiv.org/abs/0802.1189}{{\ttfamily
  0802.1189}}].

\bibitem{Cacciari:2005hq}
M.~Cacciari and G.P.~Salam, \emph{{Dispelling the $N^{3}$ myth for the $k_t$
  jet-finder}},
  \href{https://doi.org/10.1016/j.physletb.2006.08.037}{\emph{Phys. Lett. B}
  {\bfseries 641} (2006) 57}
  [\href{https://arxiv.org/abs/hep-ph/0512210}{{\ttfamily hep-ph/0512210}}].

\bibitem{NNPDF:2017mvq}
{\scshape NNPDF} collaboration, \emph{{Parton distributions from high-precision
  collider data}},
  \href{https://doi.org/10.1140/epjc/s10052-017-5199-5}{\emph{Eur. Phys. J. C}
  {\bfseries 77} (2017) 663}
  [\href{https://arxiv.org/abs/1706.00428}{{\ttfamily 1706.00428}}].

\bibitem{Buckley:2014ana}
A.~Buckley, J.~Ferrando, S.~Lloyd, K.~Nordstr\"om, B.~Page, M.~R\"ufenacht
  et~al., \emph{{LHAPDF6: parton density access in the LHC precision era}},
  \href{https://doi.org/10.1140/epjc/s10052-015-3318-8}{\emph{Eur. Phys. J. C}
  {\bfseries 75} (2015) 132} [\href{https://arxiv.org/abs/1412.7420}{{\ttfamily
  1412.7420}}].

\bibitem{PDF4LHCWorkingGroup:2022cjn}
{\scshape PDF4LHC Working Group} collaboration, \emph{{The PDF4LHC21
  combination of global PDF fits for the LHC Run III}},
  \href{https://doi.org/10.1088/1361-6471/ac7216}{\emph{J. Phys. G} {\bfseries
  49} (2022) 080501} [\href{https://arxiv.org/abs/2203.05506}{{\ttfamily
  2203.05506}}].

\bibitem{Berger:2009ep}
C.F.~Berger, Z.~Bern, L.J.~Dixon, F.~Febres~Cordero, D.~Forde, T.~Gleisberg
  et~al., \emph{{Next-to-Leading Order QCD Predictions for W+3-Jet
  Distributions at Hadron Colliders}},
  \href{https://doi.org/10.1103/PhysRevD.80.074036}{\emph{Phys. Rev. D}
  {\bfseries 80} (2009) 074036}
  [\href{https://arxiv.org/abs/0907.1984}{{\ttfamily 0907.1984}}].

\bibitem{Berger:2010zx}
C.F.~Berger, Z.~Bern, L.J.~Dixon, F.~Febres~Cordero, D.~Forde, T.~Gleisberg
  et~al., \emph{{Precise Predictions for W + 4 Jet Production at the Large
  Hadron Collider}},
  \href{https://doi.org/10.1103/PhysRevLett.106.092001}{\emph{Phys. Rev. Lett.}
  {\bfseries 106} (2011) 092001}
  [\href{https://arxiv.org/abs/1009.2338}{{\ttfamily 1009.2338}}].

\bibitem{Ita:2011wn}
H.~Ita, Z.~Bern, L.J.~Dixon, F.~Febres~Cordero, D.A.~Kosower and D.~Maitre,
  \emph{{Precise Predictions for Z + 4 Jets at Hadron Colliders}},
  \href{https://doi.org/10.1103/PhysRevD.85.031501}{\emph{Phys. Rev. D}
  {\bfseries 85} (2012) 031501}
  [\href{https://arxiv.org/abs/1108.2229}{{\ttfamily 1108.2229}}].

\bibitem{Bern:2013gka}
Z.~Bern, L.J.~Dixon, F.~Febres~Cordero, S.~H\"oche, H.~Ita, D.A.~Kosower
  et~al., \emph{{Next-to-Leading Order $W + 5$-Jet Production at the LHC}},
  \href{https://doi.org/10.1103/PhysRevD.88.014025}{\emph{Phys. Rev. D}
  {\bfseries 88} (2013) 014025}
  [\href{https://arxiv.org/abs/1304.1253}{{\ttfamily 1304.1253}}].

\bibitem{Caola:2021mhb}
F.~Caola, F.A.~Dreyer, R.W.~McDonald and G.P.~Salam, \emph{{Framing energetic
  top-quark pair production at the LHC}},
  \href{https://doi.org/10.1007/JHEP07(2021)040}{\emph{JHEP} {\bfseries 07}
  (2021) 040} [\href{https://arxiv.org/abs/2101.06068}{{\ttfamily
  2101.06068}}].

\bibitem{Fadin:1975cb}
V.S.~Fadin, E.~Kuraev and L.~Lipatov, \emph{{On the Pomeranchuk Singularity in
  Asymptotically Free Theories}},
  \href{https://doi.org/10.1016/0370-2693(75)90524-9}{\emph{Phys. Lett. B}
  {\bfseries 60} (1975) 50}.

\bibitem{Kuraev:1976ge}
E.A.~Kuraev, L.N.~Lipatov and V.S.~Fadin, \emph{Multi - {R}eggeon processes in
  the {Y}ang-{M}ills theory}, {\emph{Sov. Phys. JETP} {\bfseries 44} (1976)
  443}.

\bibitem{Kuraev:1977fs}
E.A.~Kuraev, L.N.~Lipatov and V.S.~Fadin, \emph{The {P}omeranchuk singularity
  in nonabelian gauge theories}, {\emph{Sov. Phys. JETP} {\bfseries 45} (1977)
  199}.

\bibitem{Balitsky:1978ic}
I.I.~Balitsky and L.N.~Lipatov, \emph{The {P}omeranchuk singularity in quantum
  chromodynamics}, {\emph{Sov. J. Nucl. Phys.} {\bfseries 28} (1978) 822}.

\bibitem{Andersen:2020yax}
J.R.~Andersen, J.A.~Black, H.M.~Brooks, E.P.~Byrne, A.~Maier and J.M.~Smillie,
  \emph{{Combined subleading high-energy logarithms and NLO accuracy for W
  production in association with multiple jets}},
  \href{https://doi.org/10.1007/JHEP04(2021)105}{\emph{JHEP} {\bfseries 04}
  (2021) 105} [\href{https://arxiv.org/abs/2012.10310}{{\ttfamily
  2012.10310}}].

\bibitem{Jager:2010aj}
B.~Jager, \emph{{Next-to-leading order QCD corrections to photon production via
  weak-boson fusion}},
  \href{https://doi.org/10.1103/PhysRevD.81.114016}{\emph{Phys. Rev. D}
  {\bfseries 81} (2010) 114016}
  [\href{https://arxiv.org/abs/1004.0825}{{\ttfamily 1004.0825}}].

\bibitem{Baglio:2024gyp}
J.~Baglio et~al., \emph{{Release note: VBFNLO 3.0}},
  \href{https://doi.org/10.1140/epjc/s10052-024-13336-x}{\emph{Eur. Phys. J. C}
  {\bfseries 84} (2024) 1003}
  [\href{https://arxiv.org/abs/2405.06990}{{\ttfamily 2405.06990}}].

\bibitem{Dokshitzer:1991he}
Y.L.~Dokshitzer, V.A.~Khoze and T.~Sjostrand, \emph{{Rapidity gaps in Higgs
  production}}, \href{https://doi.org/10.1016/0370-2693(92)90312-R}{\emph{Phys.
  Lett. B} {\bfseries 274} (1992) 116}.

\bibitem{ATLAS:2024png}
{\scshape ATLAS} collaboration, \emph{{Measurements of jet cross-section ratios
  in 13~TeV proton-proton collisions with ATLAS}},
  \href{https://doi.org/10.1103/PhysRevD.110.072019}{\emph{Phys. Rev. D}
  {\bfseries 110} (2024) 072019}
  [\href{https://arxiv.org/abs/2405.20206}{{\ttfamily 2405.20206}}].

\bibitem{GridPP:2006wnd}
{\scshape GridPP} collaboration, \emph{{GridPP: Development of the UK computing
  Grid for particle physics}},
  \href{https://doi.org/10.1088/0954-3899/32/1/N01}{\emph{J. Phys. G}
  {\bfseries 32} (2006) N1}.

\bibitem{Britton:2009ser}
D.~Britton et~al., \emph{{GridPP: the UK grid for particle physics}},
  \href{https://doi.org/10.1098/rsta.2009.0036}{\emph{Phil. Trans. Roy. Soc.
  Lond. A} {\bfseries 367} (2009) 2447}.

\bibitem{Andersen:2016vkp}
J.R.~Andersen, J.J.~Medley and J.M.~Smillie, \emph{{$Z/\gamma^{*}$ plus
  multiple hard jets in high energy collisions}},
  \href{https://doi.org/10.1007/JHEP05(2016)136}{\emph{JHEP} {\bfseries 05}
  (2016) 136} [\href{https://arxiv.org/abs/1603.05460}{{\ttfamily
  1603.05460}}].

\bibitem{DelDuca:1995hf}
V.~Del~Duca, \emph{{An introduction to the perturbative QCD pomeron and to jet
  physics at large rapidities}},
  \href{https://arxiv.org/abs/hep-ph/9503226}{{\ttfamily hep-ph/9503226}}.

\bibitem{DelDuca:1995zy}
V.~Del~Duca, \emph{{Equivalence of the Parke-Taylor and the
  Fadin-Kuraev-Lipatov amplitudes in the high-energy limit}},
  \href{https://doi.org/10.1103/PhysRevD.52.1527}{\emph{Phys.Rev.} {\bfseries
  D52} (1995) 1527} [\href{https://arxiv.org/abs/hep-ph/9503340}{{\ttfamily
  hep-ph/9503340}}].

\bibitem{DelDuca:1999ha}
V.~Del~Duca, A.~Frizzo and F.~Maltoni, \emph{{Factorization of tree QCD
  amplitudes in the high-energy limit and in the collinear limit}},
  \href{https://doi.org/10.1016/S0550-3213(99)00657-4}{\emph{Nucl.Phys.}
  {\bfseries B568} (2000) 211}
  [\href{https://arxiv.org/abs/hep-ph/9909464}{{\ttfamily hep-ph/9909464}}].

\bibitem{Andersen:2009nu}
J.R.~Andersen and J.M.~Smillie, \emph{{Constructing All-Order Corrections to
  Multi-Jet Rates}}, \href{https://doi.org/10.1007/JHEP01(2010)039}{\emph{JHEP}
  {\bfseries 01} (2010) 039} [\href{https://arxiv.org/abs/0908.2786}{{\ttfamily
  0908.2786}}].

\bibitem{Andersen:2009he}
J.R.~Andersen and J.M.~Smillie, \emph{{The Factorisation of the t-channel Pole
  in Quark-Gluon Scattering}},
  \href{https://doi.org/10.1103/PhysRevD.81.114021}{\emph{Phys. Rev. D}
  {\bfseries 81} (2010) 114021}
  [\href{https://arxiv.org/abs/0910.5113}{{\ttfamily 0910.5113}}].

\bibitem{Andersen:2011hs}
J.R.~Andersen and J.M.~Smillie, \emph{{Multiple Jets at the LHC with High
  Energy Jets}}, \href{https://doi.org/10.1007/JHEP06(2011)010}{\emph{JHEP}
  {\bfseries 06} (2011) 010} [\href{https://arxiv.org/abs/1101.5394}{{\ttfamily
  1101.5394}}].

\bibitem{DelDuca:2001gu}
V.~Del~Duca and E.W.N.~Glover, \emph{{The High-energy limit of QCD at two
  loops}}, \href{https://doi.org/10.1088/1126-6708/2001/10/035}{\emph{JHEP}
  {\bfseries 10} (2001) 035}
  [\href{https://arxiv.org/abs/hep-ph/0109028}{{\ttfamily hep-ph/0109028}}].

\bibitem{Bogdan:2002sr}
A.V.~Bogdan, V.~Del~Duca, V.S.~Fadin and E.W.N.~Glover, \emph{{The Quark Regge
  trajectory at two loops}},
  \href{https://doi.org/10.1088/1126-6708/2002/03/032}{\emph{JHEP} {\bfseries
  03} (2002) 032} [\href{https://arxiv.org/abs/hep-ph/0201240}{{\ttfamily
  hep-ph/0201240}}].

\bibitem{Buccioni:2024gzo}
F.~Buccioni, F.~Caola, F.~Devoto and G.~Gambuti, \emph{{Investigating the
  universality of five-point QCD scattering amplitudes at high energy}},
  \href{https://doi.org/10.1007/JHEP03(2025)129}{\emph{JHEP} {\bfseries 03}
  (2025) 129} [\href{https://arxiv.org/abs/2411.14050}{{\ttfamily
  2411.14050}}].

\bibitem{Abreu:2024xoh}
S.~Abreu, G.~De~Laurentis, G.~Falcioni, E.~Gardi, C.~Milloy and L.~Vernazza,
  \emph{{The two-loop Lipatov vertex in QCD}},
  \href{https://doi.org/10.1007/JHEP04(2025)161}{\emph{JHEP} {\bfseries 04}
  (2025) 161} [\href{https://arxiv.org/abs/2412.20578}{{\ttfamily
  2412.20578}}].

\bibitem{Maguire:2017ypu}
E.~Maguire, L.~Heinrich and G.~Watt, \emph{{HEPData: a repository for high
  energy physics data}},
  \href{https://doi.org/10.1088/1742-6596/898/10/102006}{\emph{J. Phys. Conf.
  Ser.} {\bfseries 898} (2017) 102006}
  [\href{https://arxiv.org/abs/1704.05473}{{\ttfamily 1704.05473}}].

\end{thebibliography}\endgroup

\end{document}